\documentclass[preprint,12pt]{elsarticle}
\usepackage{amsmath,amssymb}

\journal{Physica A}
\usepackage{graphicx}
\usepackage[mathscr]{eucal}
\usepackage{color}
\usepackage{multirow}
\usepackage{widetext}
\usepackage{lipsum}
\usepackage{graphicx}
\usepackage{bm}
\usepackage{dcolumn}
\usepackage[mathscr]{eucal}
\usepackage{color}
\usepackage{comment}
\usepackage{float}

\begin{document}

\begin{frontmatter}
\title{Correlations and Clustering in \\Wholesale Electricity Markets}
\author{Tianyu Cui}
\address{Science \& Technology on Integrated Information System Laboratory\\
	Institute of Software, Chinese Academy of Sciences, Beijing 100190, China}
\author{Francesco Caravelli}
\address{Theoretical Division and Center for Nonlinear Studies,\\
Los Alamos National Laboratory, Los Alamos, New Mexico 87545, USA}
\author{Cozmin Ududec}
\address{Invenia Labs, 27 Parkside Place, Parkside, Cambridge CB1 1HQ, UK}

\begin{abstract}
We study the structure of locational marginal prices in day-ahead and real-time wholesale electricity markets. In particular, we consider the case of two North American markets and show that the price correlations contain information on the locational structure of the grid.  We study various clustering methods and introduce a type of correlation function based on event synchronization for spiky time series, and another based on string correlations of location names provided by the markets. This allows us to reconstruct aspects of the locational structure of the grid.

\end{abstract}


\end{frontmatter}


\section{Introduction}

Electricity is different from other commodities in at least three important respects: it cannot currently be stored efficiently, it flows through the electricity grid according to the laws of electromagnetism rather than directly from a producer to a consumer, and at any moment supply and demand must match almost exactly to avoid blackouts or other issues.
The capacity of the transmission network also limits the amounts of power that can be injected or withdrawn at certain locations. Therefore, in order for an electricity grid to function safely and efficiently, the configuration and physical limits of the grid and the locations of generation and consumption must be taken into account when making decisions about consumption and production.  To this end, many grid operators around the world have adopted a spatial and temporal pricing mechanism known as Locational Marginal Pricing (LMP) \cite{Bohn}.  This mechanism sets potentially different prices at hundreds or thousands of important locations (nodes) throughout an electricity grid.  These prices are part of the solution to an Optimal Power Flow (OPF) problem \cite{OPF}, and represent the increase in optimized total system costs as a result of a small increase in the amount of power consumed at a location at a specific time. 

In this paper we study the properties of LMPs in two North American wholesale electricity markets: Pennsylvania-New Jersey-Maryland (PJM) \cite{PJM}, and Midcontinent Independent System Operator (MISO) \cite{MISO}.  In particular, in both of these markets there are two types of LMPs: day-ahead (forward) LMPs, and real-time (spot) LMPs \cite{ferc1}.  The day-ahead prices are outputs of what can be thought of as a planning exercise by the system operator, in which OPF problems are run every day for each hour of the next (target day).  The real-time LMPs are then the outputs of the OPF problems that are solved in real time (usually every 5 or 15 minutes) to adjust the schedules output from the day-ahead runs.

Our main goal is to understand what information can be inferred from day-ahead and real-time LMPs.  Specifically, we are interested in whether any underlying structure can be inferred from the correlations between prices. As discussed above, LMPs result from a highly structured and constrained optimization problem \cite{Zimmerman}, which induces non-local correlations between prices. It is therefore interesting to understand how prices are correlated according to various measures, how clusters of nodes emerge, and whether these can be understood in terms of the spatial distribution of nodes in the grid.


We will use a combination of methods introduced in the machine learning and statistical physics literatures. 
One interesting approach begins by first \textit{filtering} the relevant information from covariance and correlation matrices.  This approach has mainly been used in the financial literature, in order to remove possible noise and to focus only on the most important structures.  We will have to introduce a few caveats and differences which are due to the underlying physical structure of power grids \cite{potters1, aste1, Caravelli}. Given the specific spiky nature of the time series, as an alternative to the Pearson correlation, we also introduce a method commonly used for neural networks, \textit{event synchronization}, to measure the synchronization between spikes at various locations. 
We will also test another correlation measure introduced in machine learning: the graphical lasso method. This is then used to cluster the time series.  For this purpose, we use both the Minimal Spanning Tree method, the Planar Maximally Filtered Graph algorithms approach and Spectral Clustering. We then compare the results to clustering based on the node names as a proxy for node location.

\section{Data structure}\label{sec:data}

The dataset we will analyse has been collected by Invenia TCC. A subset of the data was used in \cite{Caravelli}, where a description of the data is also provided \cite{Data}.  The time series we consider is available for each node in the PJM and MISO grids for a period of 3 years (January 1, 2012 to December 31, 2015) at hourly resolution. We focus on a subset of 1287 nodes for PJM and 2568 nodes in MISO (these are nodes which virtual participants can transact on).

The full (day-ahead or real-time) price of electricity at node $n$ and  time $t$ is given by:
\begin{equation}
LMP(n,t) = MEC(t) + MCC(n,t) + MLC(n,t),
\end{equation}
where MEC stands for \textit{Marginal Energy Cost}, MCC stands for \textit{Marginal Congestion Cost} and MLC stands for \textit{Marginal Loss Cost}. 
The $MCC(n,t)$ component is the price due to transmission congestion, i.e., it is the marginal cost of supplying the next increment of load at a location, taking into account the transmission constraints of the grid.  This can be positive or negative, and is often $0$.  For example, when a power line at some location is at its limit for carrying power, the load at that location must be serviced through another line, which can be more costly.
$MLC(n,t)$ is the price due transmission losses on the grid.  This is generally small compared to $MEC$.
$MEC(t)$ can roughly be thought of as the price of electricity at any given node if there is no congestion and loss to that node.
The $MEC$ component is independent of node, and thus represents a price shift for the whole market. 
The $MLC$ and $MCC$ components are instead time and node dependent, but we observe that in general $MCC\gg MLC$.
Thus, for the present paper we study only the MCC component of the prices in the day ahead market, which directly address the inefficiency of power transmission and is the main source of volatility in the $LMP$ time series.  An important quantity for many market participants and for the grid operator is the difference between day-ahead and real-time prices:
\begin{equation}
\Delta(n,t)=DA(n,t)-RT(n,t).
\end{equation}
This is the time series we will focus on. 

One important detail which will be important for our analysis is the fact that each nodal price has a string associated to it, which is of the form ``NODENAME\_CODE". The text ``NODENAME" can be loosely associated with the location of the node. We can in fact run a clustering algorithm based on the name and compare to the results based on price correlations only. This will allow us to compare correlation matrices obtained from independent methods and identify the best price correlation technique to pair nodes and associated clusters.

\section{Correlations, synchronization measures, and filtering processes}
We begin by introducing several measures of interdependence between time series which will be relevant in our analysis.

\subsection{Pearson Correlation}
The Pearson correlation is one of the most commonly used and simplest measures.  It assumes stationarity and a linear relationship between the time series. The correlation matrix is defined as:
\begin{equation}
C_{ij}=Corr[X_i,X_j]= \frac{Cov[X_i,X_j]}{\sqrt{Var[X_i] Var[X_j]}},
\end{equation}
where
\begin{equation}
Cov[X_i,X_j]= \langle X_i X_j\rangle - \langle X_i\rangle \langle X_j\rangle,
\end{equation}
and $Var[X_i]=\sigma_i^2=\langle X_i^2\rangle-\langle X_i\rangle^2$. The averages are defined as temporal ones, e.g. $\langle X_i\rangle=T^{-1} \sum_{t=1}^T x_i(t)$.

The limitations of the Pearson correlation are well know, but we will nevertheless use it as a benchmark in what follows. The Pearson correlation matrix is shown in the first row of Fig. \ref{fig:Pearson, exp, es} for MISO (left) and  PJM (right).

\begin{figure}
	\includegraphics[scale=0.45]{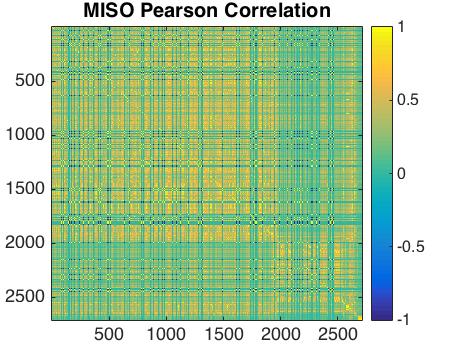}
	\includegraphics[scale=0.45]{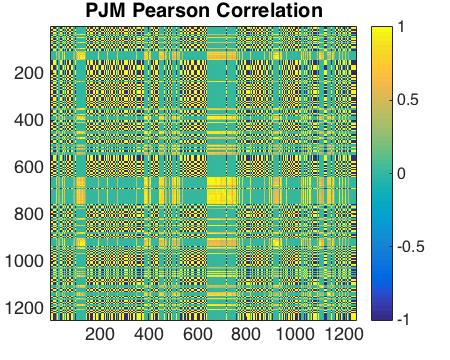}\\
	\includegraphics[scale=0.45]{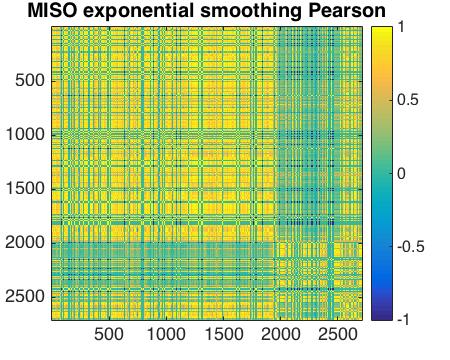}
	\includegraphics[scale=0.45]{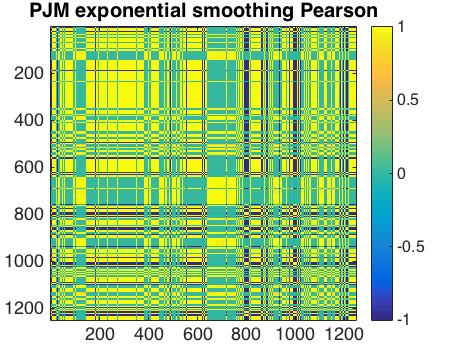}\\
	\includegraphics[scale=0.45]{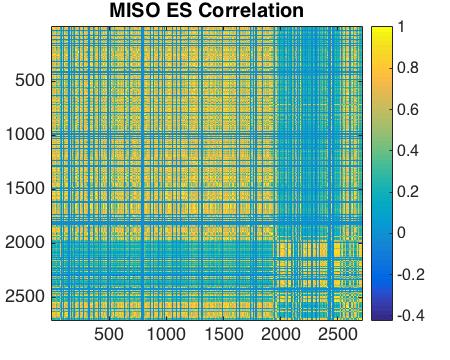}
	\includegraphics[scale=0.45]{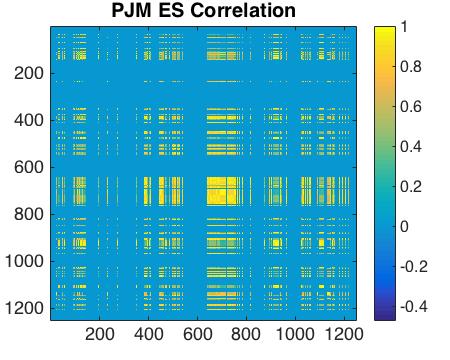}
	\caption{\textbf{Pearson correlation, exponential smoothing Pearson and event synchronization} for MISO (left) and PJM (right) market.}
	\label{fig:Pearson, exp, es}
\end{figure}

\subsection{Exponential smoothing for Pearson correlation}
Assigning the same weight to each time point (standard Pearson correlation) assumes that recent events and remote past events have the same importance, which is often not a realistic assumption for many purposes \cite{pozzi2012exponential,musmeci2016multiplex}.
In order to take this idea into account, we need to assign different weights for different time points, and the weights should satisfy the following three requirements: $(i)$ $\mathbf{w}\geq0$, $(ii)$ $\forall u > v, w_{u} > w_{v}$, and $(iii)$ $\sum_{t=1}^{T}w_{t}=1$.  The weighted Pearson correlation is then defined as
	\begin{equation}
	\varrho_{ij}^{w} = \frac{\sigma_{ij}^{w}}{\sigma_{i}^{w}\sigma_{j}^{w}},
	\end{equation}
	where $\sigma_{k}^{w}=\sqrt{\sum_{t=1}^{T}w_{t}(X_{kt} - \hat{X_{k}})^{2}}$, $\sigma_{ij}^{w} = \sum_{t = 1}^{T}w_{t}(X_{it} - \hat{X_{i}})(X_{jt} - \hat{X_{j}})$, and $\hat{X_{k}} = \sum_{t = 1}^{T}w_{t}X_{kt}$. 

A common way to construct the weights vector is by using an exponential function which leads to the  \textit{exponentially smoothed Pearson correlation}, where
\begin{equation}
w_{t} = w_{0}\exp\left(\frac{t - T}{\theta}\right),
\label{exponential smoothing}
\end{equation}
where $t\in{1, 2, 3,\ldots,T}$ and $\theta > 0$ is the weights decay factor. We can control the smoothness of the weights by varying $\theta$. If $\theta \rightarrow 0$, only the most recent data point is relevant, and if $\theta \rightarrow \infty$, the smoothed correlation converges to the Pearson correlation. 
The experimental results for $\theta=3$ are shown in the second row of Fig. \ref{fig:Pearson, exp, es} for MISO (left) and  PJM (right).

\subsection{Event Synchronization}

\begin{figure}
\centering
\includegraphics[scale=0.43]{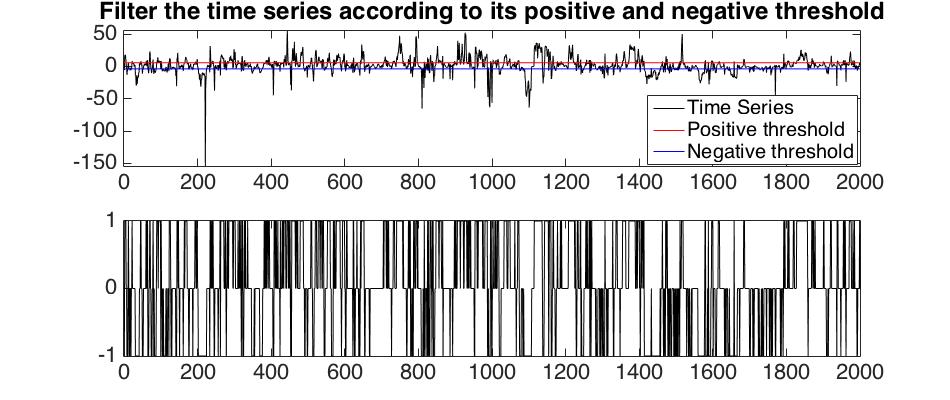}
\caption{\textbf{Illustration of event synchronization.} \textit{(top)} Filtering the MCC time series with extremely spiky behavior according to its positive and negative median value.  \textit{(bottom)} Filtered time series obtained by thresholding.}
\label{fig:filt}
\end{figure}

Given the spiky nature of electricity prices, we consider another distance between time series which is a variation of event synchronization, which is inspired by other studies of spiky time series, such EEG time series \cite{Kreuz}. A spike in the MCC time series can be intuitively associated to the failure of efficient dispatch and congestion in the power grid (prices are related to Lagrange multipliers of constraints that are imposed on the system \cite{Tesfatsion}). 
We introduce the quantity $c^{\tau}(x|y)$, which is the number of times an event appears in $x$ shortly after it appears in $y$,
\begin{equation}
c^\tau(x|y)=\sum_i \sum_j J^\tau_{ij},
\end{equation}
where $J^\tau_{ij}=1$ if the spikes occur within the interval $\tau$, i.e. $0<|t_i^x-t_i^y|\le \tau$, $J_{ij}=1/2$ if $t_i^x=t_i^y$ and zero otherwise.
We then define 
$Q_{\tau}=\frac{c^\tau(y|x)+c^\tau(x|y)}{\sqrt{m_x m_y}}$
which is a symmetrized matrix based on the counting above.
$Q_\tau$ is not a measure of correlation yet, as it has to be normalized. Further note that this measure does not consider the fact that two spikes can be anti-correlated. 

Here we generalize the above by first introducing a notion of ``event".
Given a time series $x(t)$ with values which can be positive or negative, we introduce the positive and negative thresholds $mp_{x}$ as the median of the positive values, and $mn_{x}$ as the median of the negative values of $x$ respectively, calculated in a time window of length $T$. 
We consider a positive event as a particular time where $x(t)$ is above the threshold $mp_{x}$, and a negative event as one where $x(t)$ below $mn_{x}$. 
When the time series is between $mp_{x}$ and $mn_{x}$ the filtered signal is set to zero. 
This results in a transformed time series $\epsilon_x(t) \in \{-1, 0, 1\}$.
This filtering is depicted in Fig. \ref{fig:filt}.

\begin{figure}
	\includegraphics[scale=0.45]{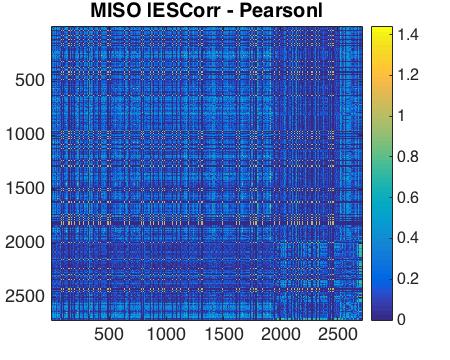}
	\includegraphics[scale=0.45]{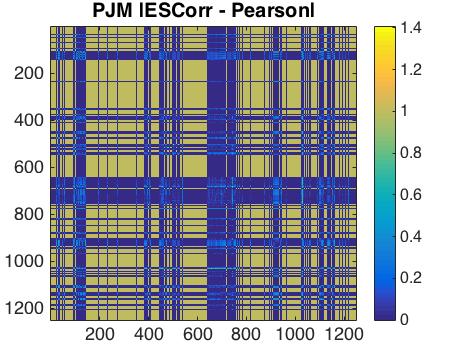}\\
	\caption{\textbf{Absolute value of the difference between standard Pearson correlation and Modified Event Synchronization}, for MISO (left) and PJM (right).}
	\label{fig:escorr}
\end{figure}

We can then construct the matrix $J^{\tau}_{ij}$ as follows:
\begin{equation}
J_{ij}^\tau=\sum_{t} \epsilon_{x_i}(t) \sum_{t^\prime, |t-t^\prime|\leq \tau} \epsilon_{x_j}(t^\prime).
\end{equation}
We then consider the matrix $D_{ij}=\delta_{ij} \sum_{j^\prime}J_{ij^\prime}$, and normalize the synchronization matrix $J_{ij}^\tau$, obtaining an alternative correlation matrix:
\begin{equation}
C^\prime=\sqrt{D^{-1}} J \sqrt{D^{-1}},
\end{equation}
which is symmetric by construction. This measure for $\tau=3$ h is shown in the last row of Fig. \ref{fig:Pearson, exp, es} for MISO (left) and PJM (right). The difference between the Event Synchronization correlation and the Pearson correlation is shown in Fig. \ref{fig:escorr}.  Note how the two methods obtain different results for MISO (left) and PJM (right), and that this difference is in particular visibly stronger in the case of PJM. Note that two time series can be positively correlated according to the Pearson correlation, and yet have negative correlation in terms of spikes.

\subsection{Filtered correlations using random matrix theory}

In this section we discuss and apply a technique based on Random Matrix Theory (RMT) which is widely
used to identify the non-random components of empirical Pearson correlation matrices \cite{Potters}.  
The eigenvalue of a correlation matrix based on $N$ completely random time series of
duration $T$ follows the Marchenko-Pastur (or Sengupta-Mitra) distribution:
\begin{equation}
\rho(\lambda) =\frac{1}{\rho} \frac{\sqrt{(\lambda_{+}-\lambda)(\lambda-\lambda_{-})}}{2\pi\lambda},
\end{equation}
and
\begin{equation}
\lambda_{\pm}=(1 \pm \sqrt{\rho})^2,
\end{equation}
where $\rho=\frac{T}{N}$, and $1 < \frac{T}{N} < \infty$, when $N\rightarrow \infty$ and $T\rightarrow \infty$. 

\begin{figure}[h]
\includegraphics[scale=0.45]{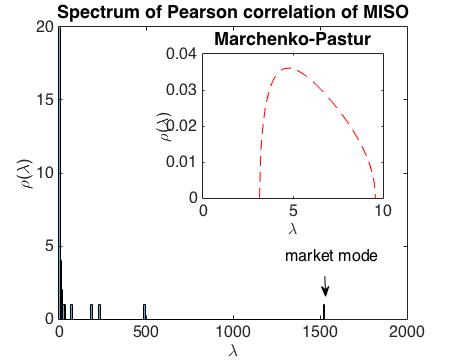}
\includegraphics[scale=0.45]{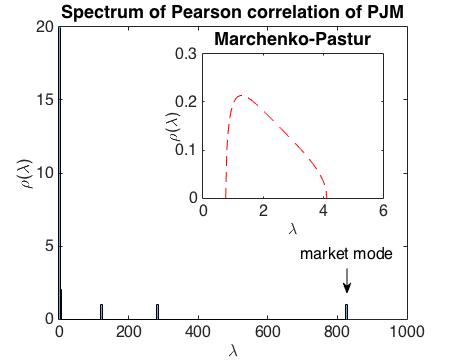}
\caption{\textbf{Pearson Correlation matrix spectrum distributions} and their corresponding \textbf{Marchenko-Pastur distributions} (inset) for MISO (left) and PJM (right). }
\label{fig:correlationmp}
\end{figure}

In general the correlation matrix can be decomposed into two parts, $C=C^r+C^s$, where $C^r$ is the ``random" component and $C^s$ is the matrix containing information.
Due to the information contained in the time series, the distribution of the spectrum of empirical correlation matrices is different from the corresponding Marchenko-Pastur distribution. This suggests filtering Gaussian noise efficiently by removing the random component given by RMT.
For instance,
in Fig. \ref{fig:correlationmp} we show the eigenvalue density of the empirical correlation matrix obtained for MISO and PJM. In the inset figure, we give the Marchenko-Pastur distribution
with the same values of N and T. 
It has been shown in many studies that a typical feature
of the spectrum of empirical correlation matrices is that the largest observed eigenvalue $M$ (denoted the market mode in Fig. \ref{fig:correlationmp} for PJM and MISO) is much larger than all other eigenvalues.
And the corresponding eigenvector has all positive elements and one can therefore
identify this eigencomponent of the correlation matrix as the so-called market mode \cite{Dong2005,MacMahon}, which is a common factor in all of the time series.

\begin{figure*}[h]
	\includegraphics[scale=0.45]{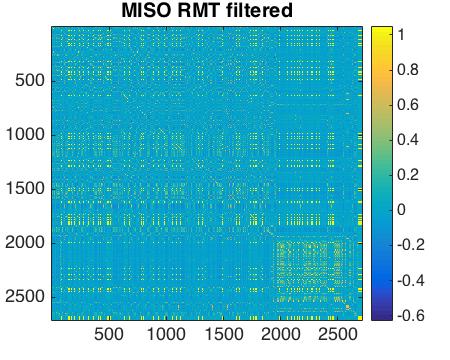} \includegraphics[scale=0.45]{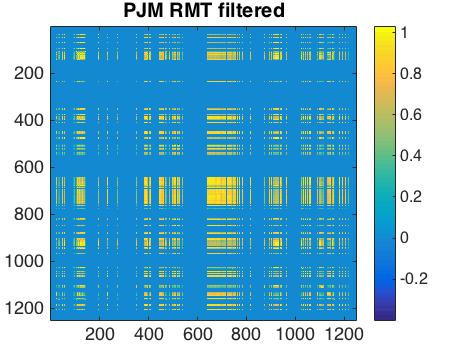}\\
	\caption{\textbf{Filtered correlations evaluated using RMT} for MISO (left) and PJM (right).}
	\label{fig:filteredRMT}
\end{figure*}

If we also consider the market component, we have the splitting
\begin{equation}
C=C^r+C^g+C^m.
\end{equation}
where $C^g$ represents correlations at the level of sub-groups of time series, which is often referred to as the ``group" mode \cite{MacMahon}, and $C^m$ represents the ``market'' component.
The result of the filtering procedure is shown in Fig. \ref{fig:filteredRMT}. In MISO we observe that most of the anti-correlation between the upper and lower block disappears, and that the lower block correlation is more pronounced. For PJM, we observe that most of the information is indeed filtered, leaving only a few strong correlation blocks.

\subsection{Graphical lasso}
The next measure of correlation we consider is Graphical Lasso, which has been introduced in statistical learning \cite{Friedman2008gl}.
This measure has several advantages over the standard Pearson correlation, especially for high dimensional time-series which live in a relatively smaller dimensional subspace. The statistical uncertainty of the Pearson correlation for such data is extremely high. This method can also be thought of as learning a pairwise Gaussian Markov Random Field (GMRF) \cite{Rue2005GMRF}, where each vertex stands for a variable and each edge exists if the corresponding variables are correlated. For empirical data, the GMRF should be a fully connected graph, because it is unlikely for two time series to have zero empirical correlation. 
The goal is then to reveal the inherent sparse graph structure from empirical data by eliminating ``weak'' edges.  
An intuitive way to solve the problem above is minimizing the reconstruction error between the Pearson correlation and the desired correlation, and shrinking the insignificant correlations to zero. This can be done  by introducing an $L_{1}$ penalty, and minimizing the functional:
\begin{equation}
\tilde{\Sigma}=\arg\min_{\Sigma}|\Sigma-\Sigma_{n}|_{F}^{2}+\rho|\Sigma|_{1},
\end{equation}
where $\Sigma_{n}$ is the empirical correlation matrix, $|\cdot|_F$ is the Frobenius norm which measures the reconstruction error, and $|\cdot|_1$ is an $L_1$ norm regularizer, with the constraint that $\Sigma$ is positive definite. Ying \cite{cui2016sparse} proposed a method to solve this optimization problem efficiently and the related algorithm can be found in references.

\begin{figure*}[h]
	\includegraphics[scale=0.45]{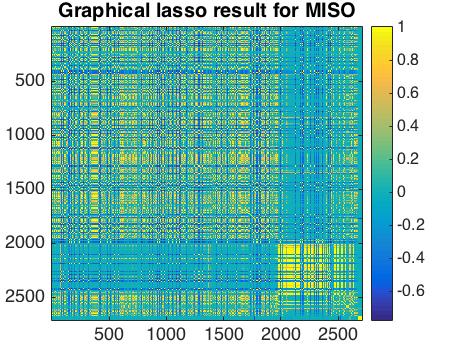} \includegraphics[scale=0.45]{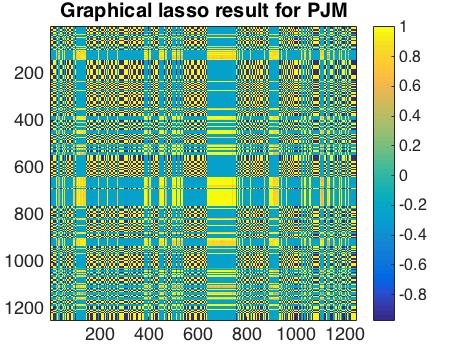}\\
	\caption{\textbf{Sparse Correlations evaluated using  graphical lasso optimization} for MISO (left) and PJM (right). }
	\label{fig:sparse}
\end{figure*}

In Fig. \ref{fig:sparse} we can see that the graphical lasso has a similar structure to the Pearson correlation shown in Fig. \ref{fig:Pearson, exp, es}. 
In a sense however this correlation matrix is more significant than Pearson (especially the lower block for MISO), because it removed many negligible elements. 

\subsection{String Correlation}\label{sec: string}
Next we introduce a correlation measure which is based on the names of the LMP time series. We will use this as an independent method to test the validity of the preceding correlations measures. In the two electricity grids we are studying each LMP time series is assigned an identifier which is given by a string of the form `place-name code-number'. This allows us to use the \textit{string kernel} \cite{kernel3, stringkernel} method to calculate the similarity between identifier strings. In the statistical learning community there are several kernel methods which are used to  estimate relationships between various types of datasets. For example, it is  possible to construct kernels for estimating the correlation between images and sentences \cite{kernel1,kernel2,kernel3}. Intuitively, a higher string kernel similarity suggests that the two time series locations are more likely to be near each other geographically.

\section{Clustering methods}

\subsection{Spectral clustering}
A standard approach for inferring structure in high dimensional time series data is to \textit{cluster} the data \cite{aste1}.
As a first example, consider the Girvan-Newman algorithm \cite{Girvan}, which infers structure by progressively removing edges from an original graph (inferred from the correlations between the time series for example), with the final remaining connected components representing `communities' in the data. 
One drawback of the Girvan-Newman algorithm is its inefficiency for large graphs, which results from the need to calculate the eigenvectors of the corresponding graph Laplacian at each iteration. Also, as noted in \cite{MacMahon}, the Girvan-Newman algorithm is not well suited for correlation matrices and requires a slight modification.

In the following we will instead use \textit{spectral clustering} which is based on the same intuition as the Girvan-Newman algorithm \cite{Von2007tutorialonSL}, but we only need to find the spectrum of the original graph Laplacian and then use the k-means algorithm to find the underlying communities.
In Fig. \ref{fig: Spectral clustering result} we show a selected sample of the original clustering results in the case of a small set of MISO nodes. The matrix elements lower than median correlation have been neglected.  A total of $200$ clusters were used in original result (see Section 5 for further discussion).

\subsection{Community detection algorithms}
Another approach to clustering is to use community detection algorithms. In order to improve the efficiency of these algorithms and filter out any noise that may be present, we will use various filtering algorithms before we cluster.
The first approach to filtering empirical correlation matrices is based on Minimal Spanning Trees (MST) which are tree structure graphs which only contain the largest correlations \cite{aste1,MacMahon,Tumminello,Gomez}. 
For a system with $N$ elements, a tree structure graph with only $N-1$ links are retained by the MST filtering algorithm. This approach is closely related to the Single Linkage hierarchical clustering algorithm \cite{musmeci2015relation}, which automatically produces an agglomerative algorithm that ends up with a dendrogram that arranges the elements into a hierarchical structure
from the original correlation matrix \cite{MacMahon, musmeci2015relation}. Previous research \cite{MacMahon} also shows that MST is more reliable when first using the strongest correlations to determine the low-level structure of the taxonomic tree, while it is progressively less reliable when using the weaker correlations to determine a high-level taxonomic tree.
A generalization of MST which releases the topological constraint to contain a larger number of links is called Planar Maximally Filtered Graph (PMFG). This retains the MST as well as a number of additional links, provided that the resulting graph is still a planar graph\footnote{The graph can be drawn on a plane without any crossing of links} \cite{MacMahon}. Since MST is associated with hierarchical clustering algorithms, PMFG has been shown to be related to Directed Bubble Hierarchical Tree (DBHT), which contain more information \cite{musmeci2015relation}.

\begin{figure*}
	\includegraphics[scale=0.125]{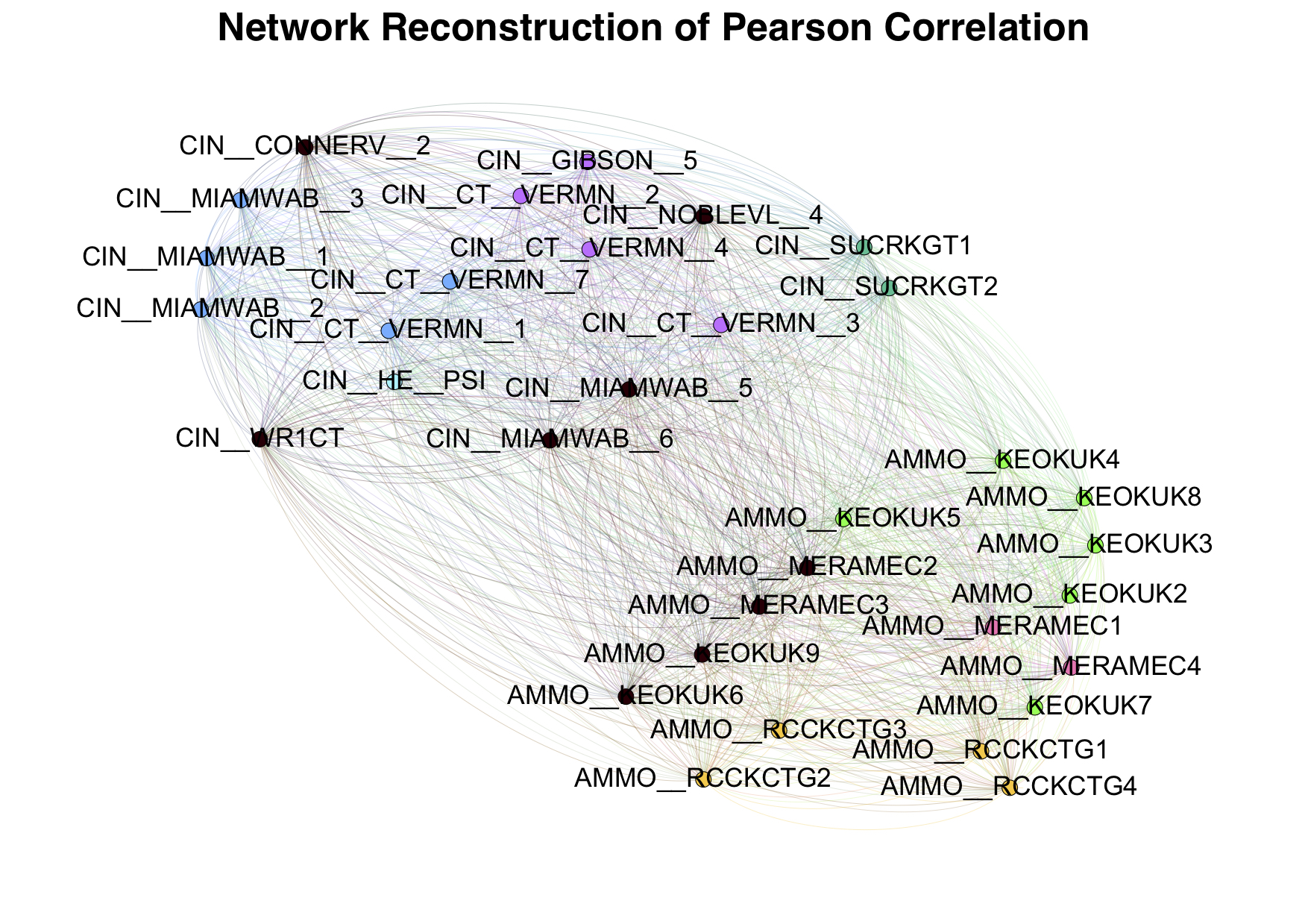}
	\includegraphics[scale=0.125]{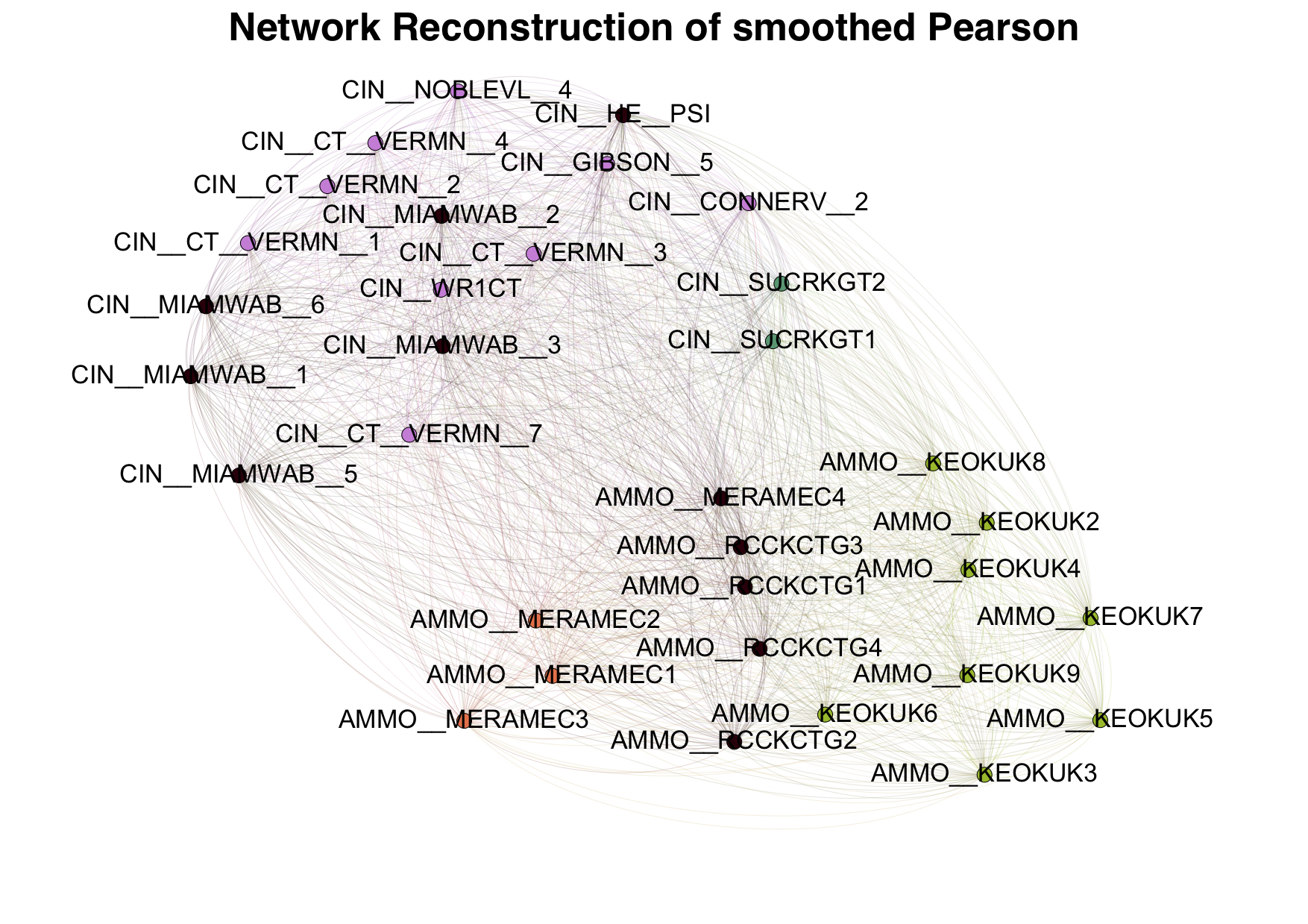}\\
	\includegraphics[scale=0.125]{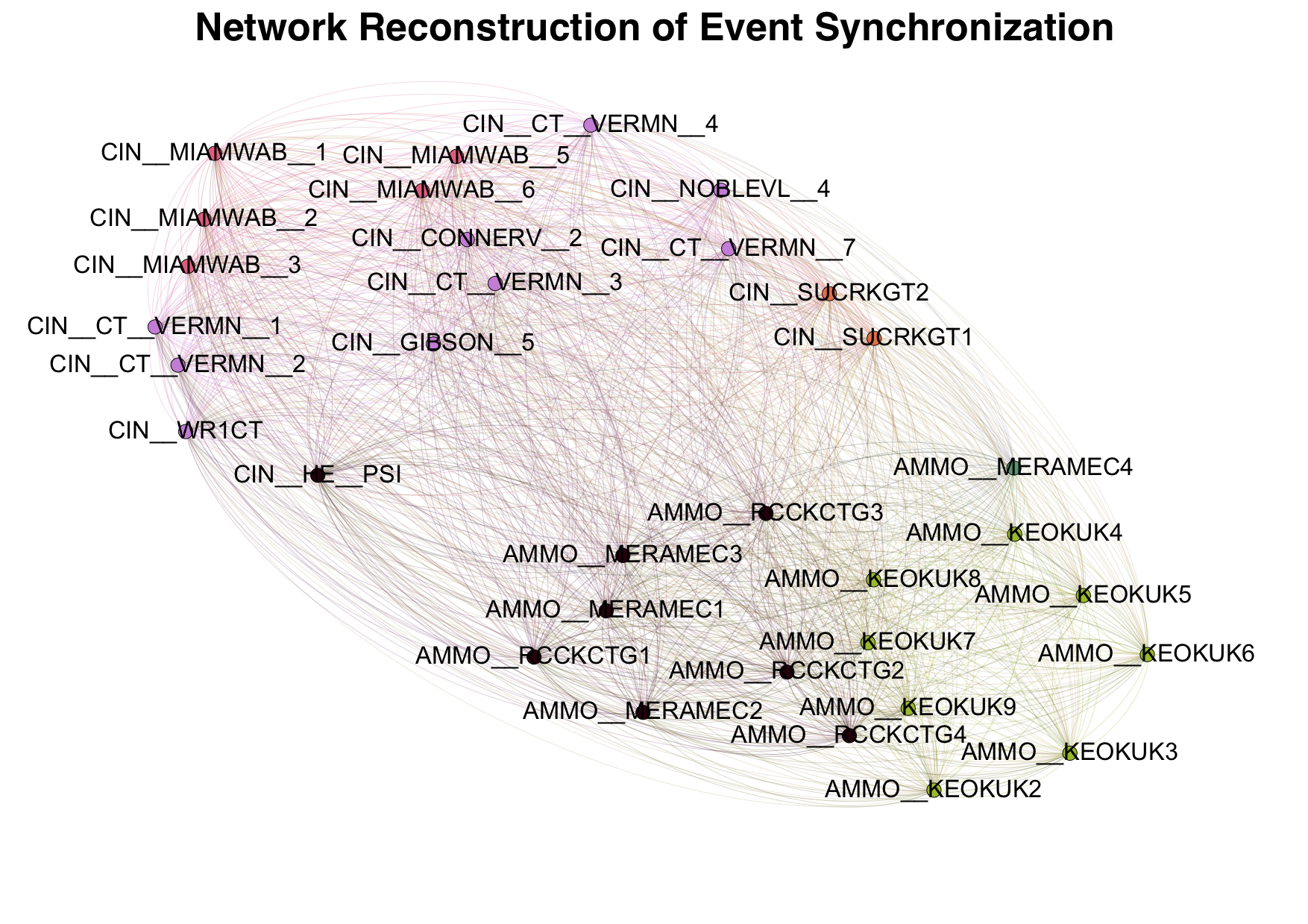}
	\includegraphics[scale=0.125]{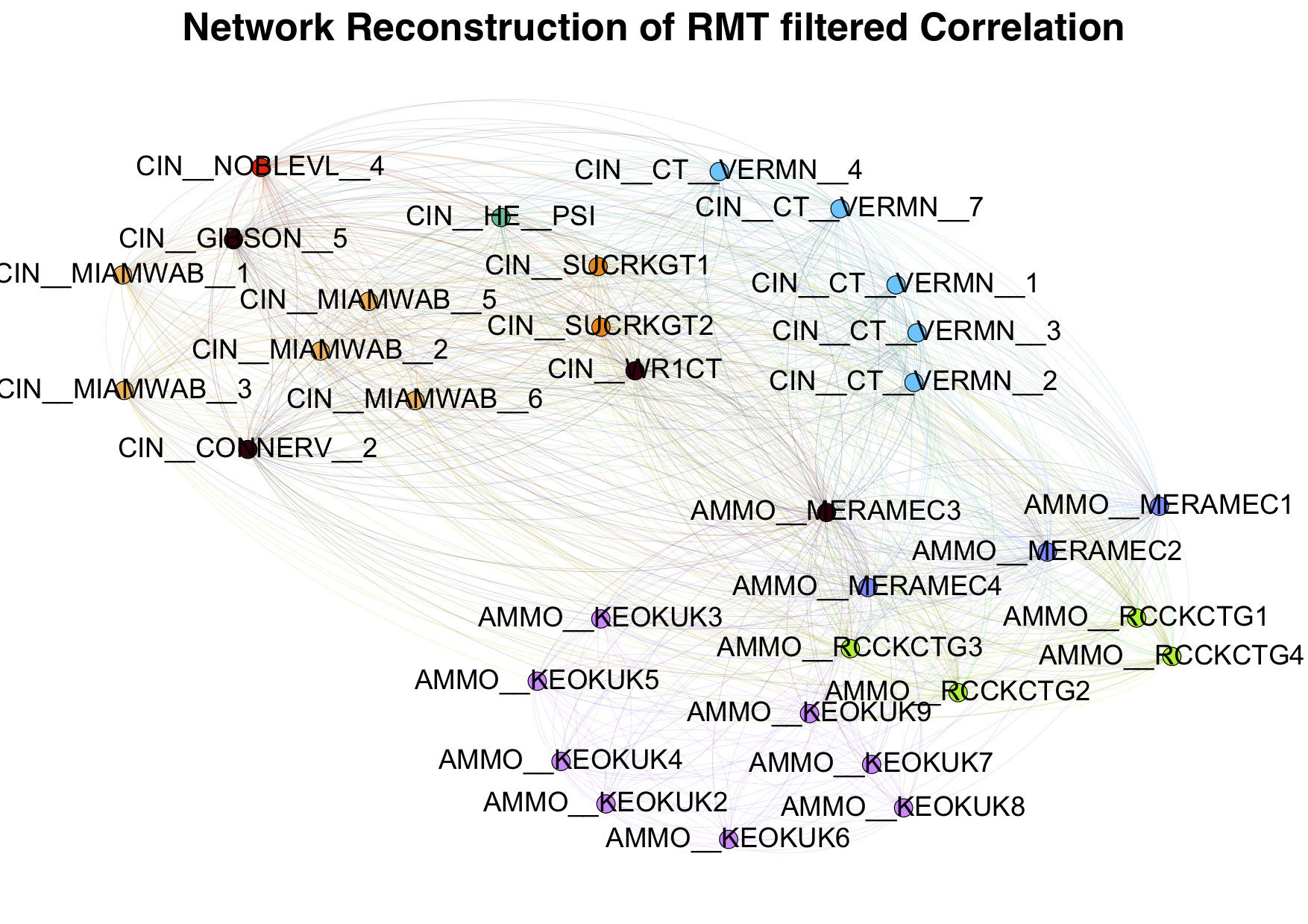}\\
	\includegraphics[scale=0.125]{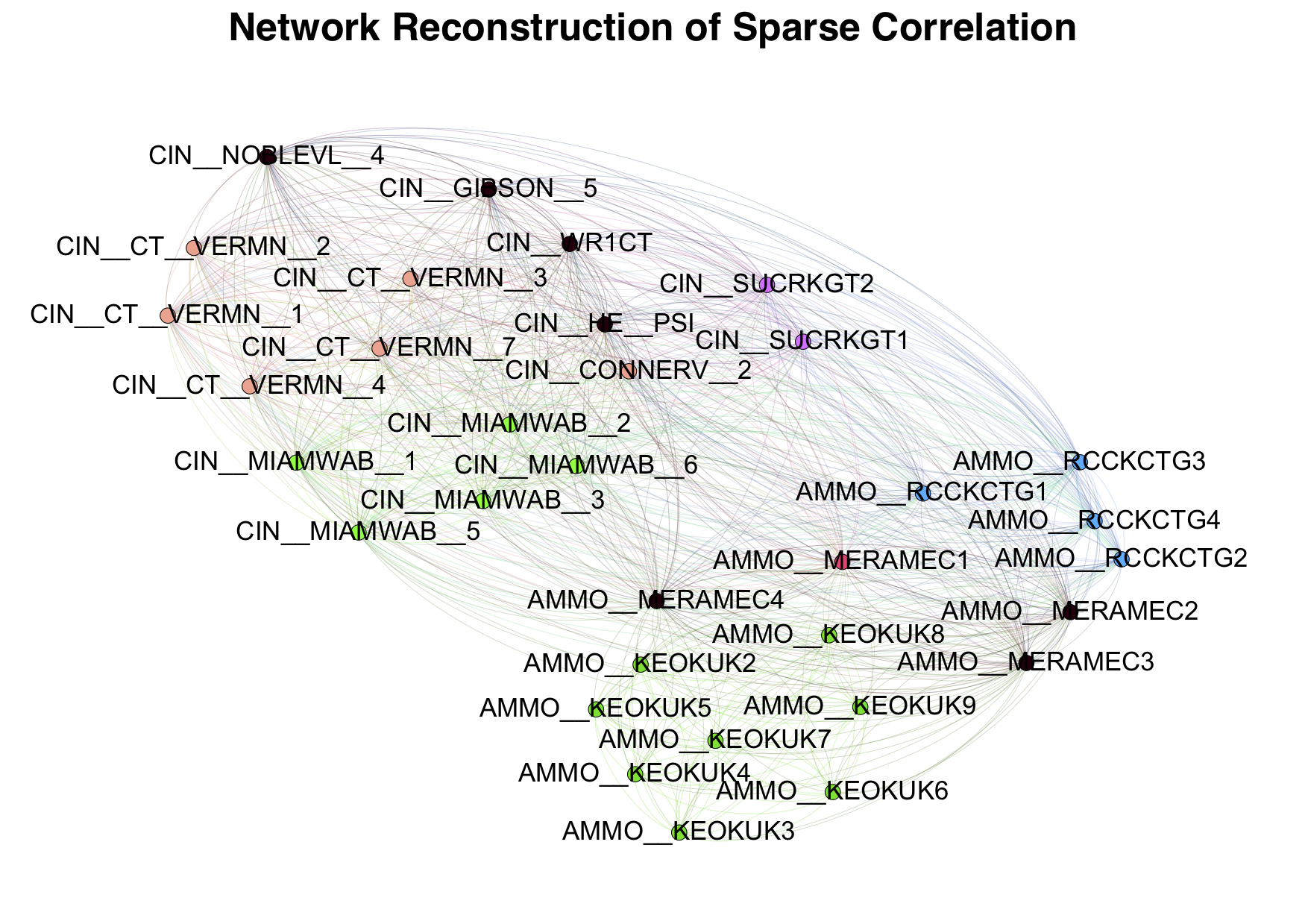}
	\includegraphics[scale=0.125]{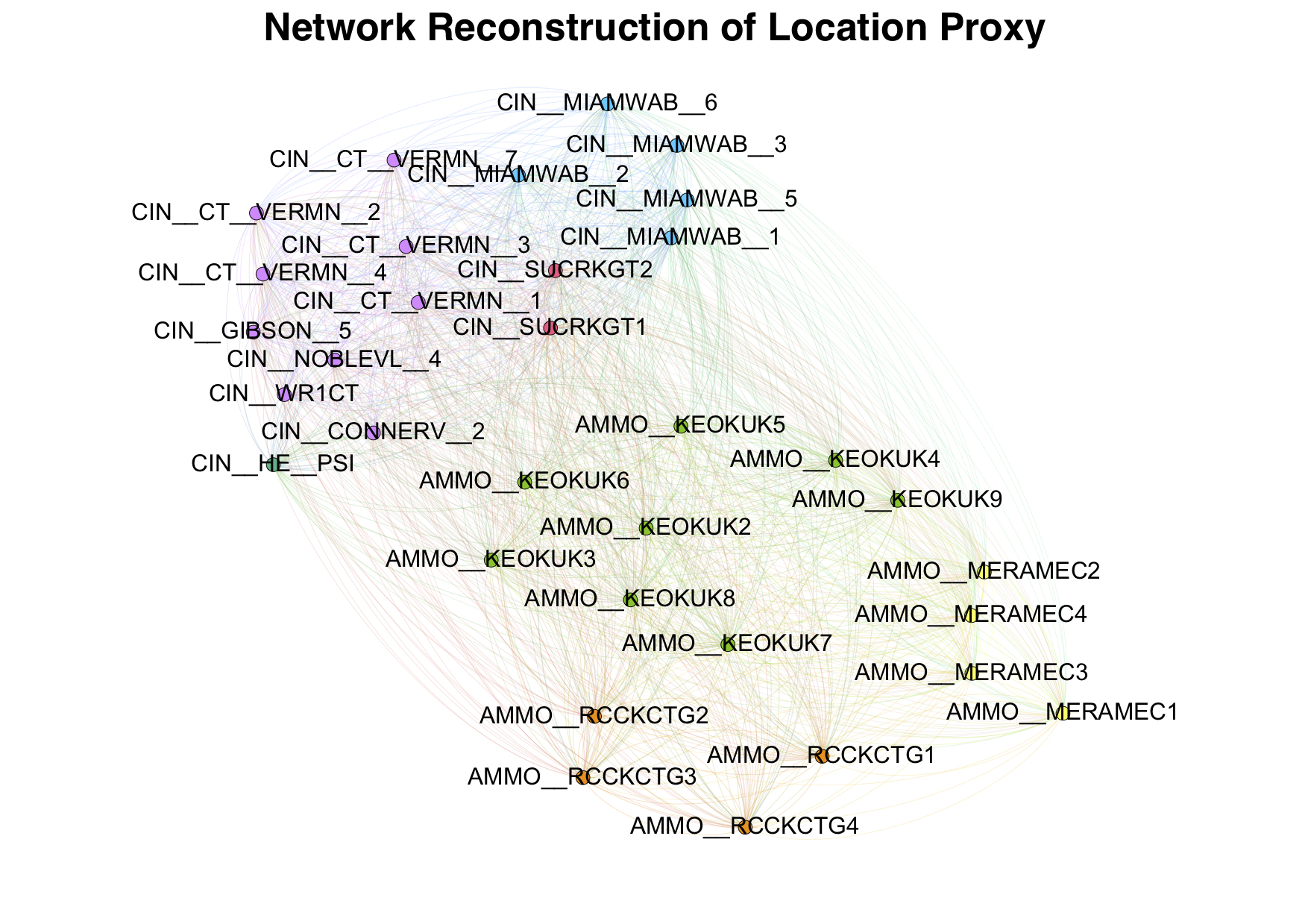}\\
	\caption{\textbf{Network reconstruction} for a small set of nodes of MISO. The location string of each node is shown in the figure. Nodes with the same color belong to the same cluster, which has been calculated by \textbf{spectral clustering}. Black nodes are those misclassified with respect to location proxy partition.  }
	\label{fig: Spectral clustering result}
\end{figure*}

\begin{figure*}
	\includegraphics[scale=0.125]{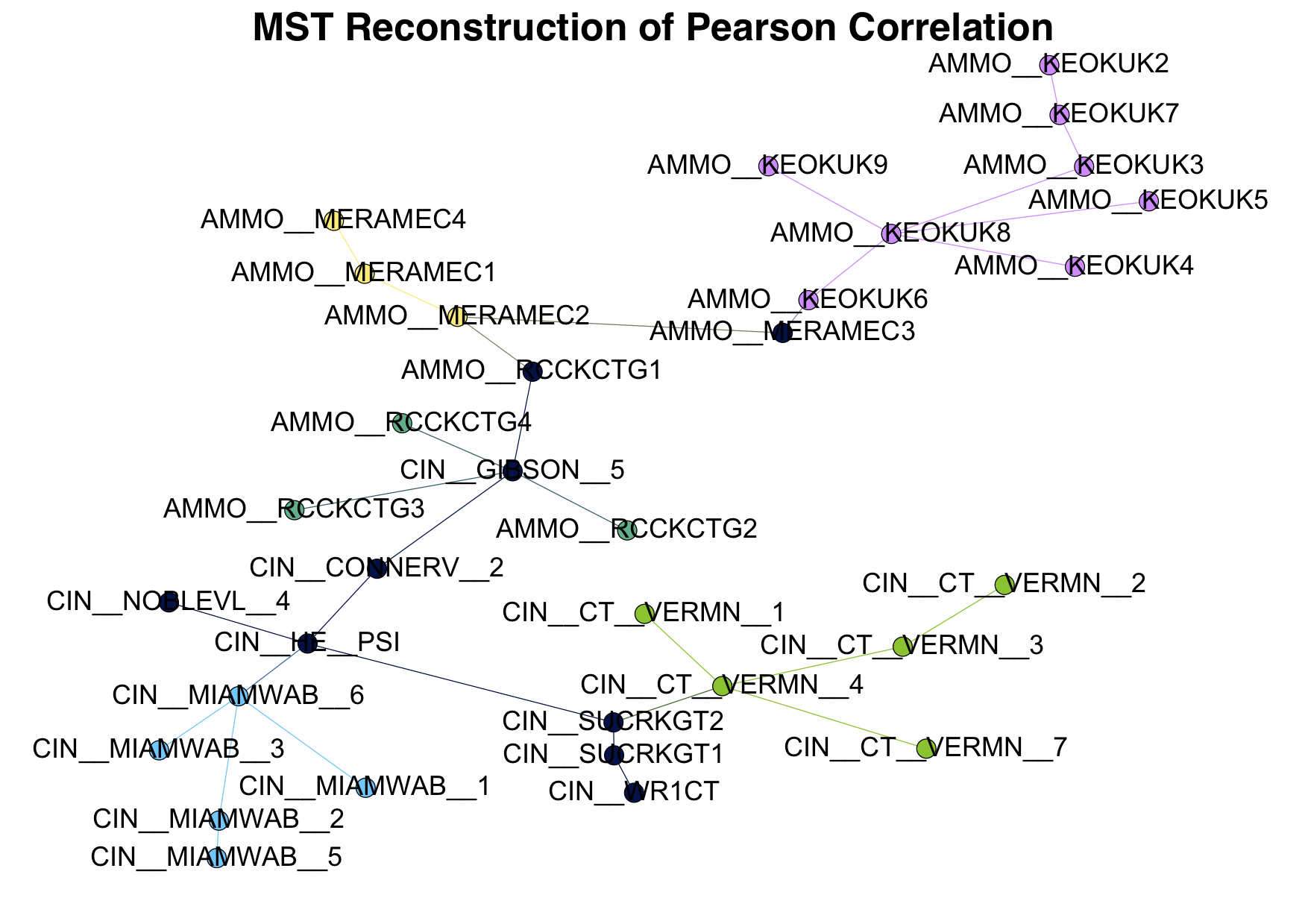}
	\includegraphics[scale=0.125]{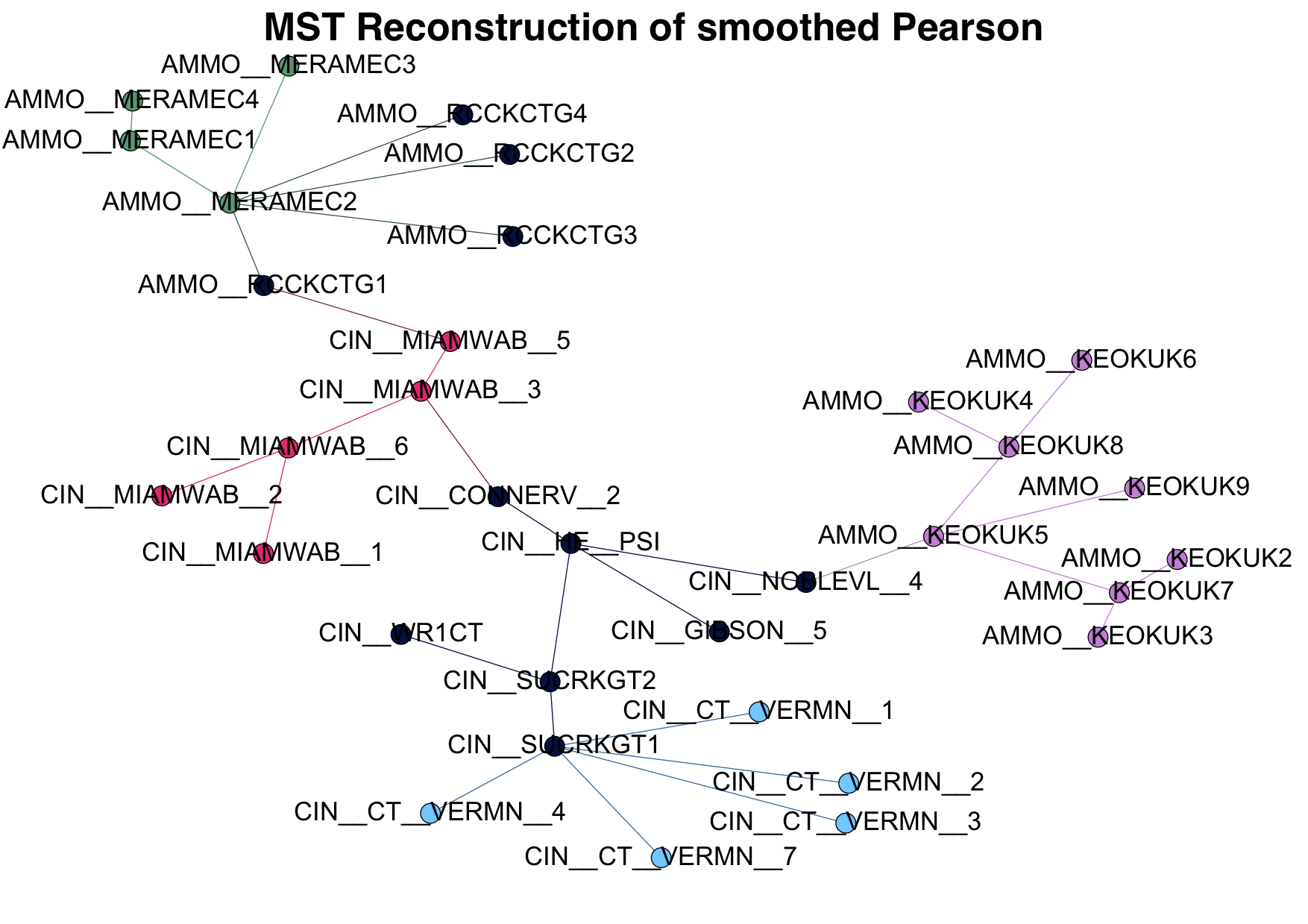}\\
	\includegraphics[scale=0.125]{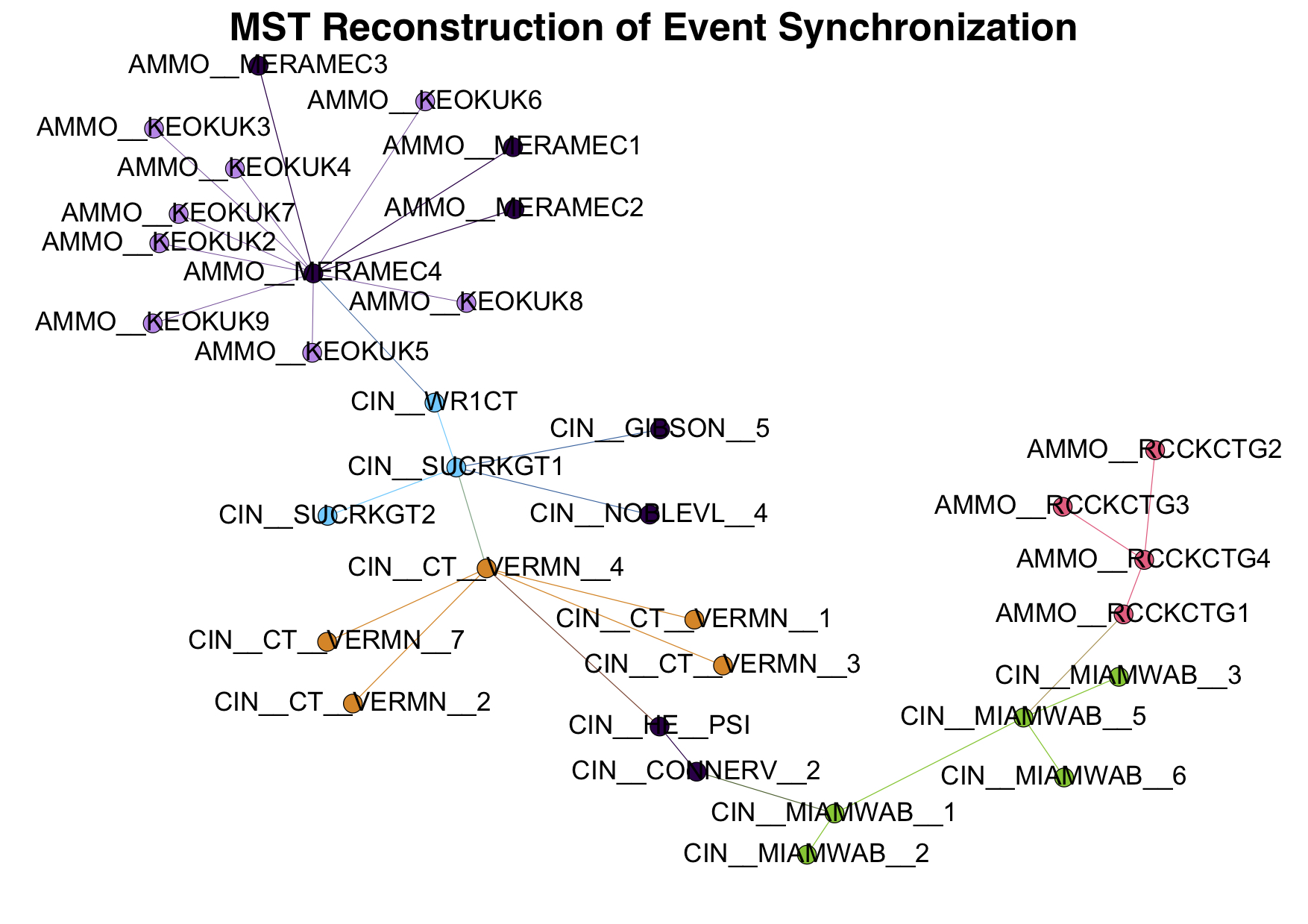}
	\includegraphics[scale=0.125]{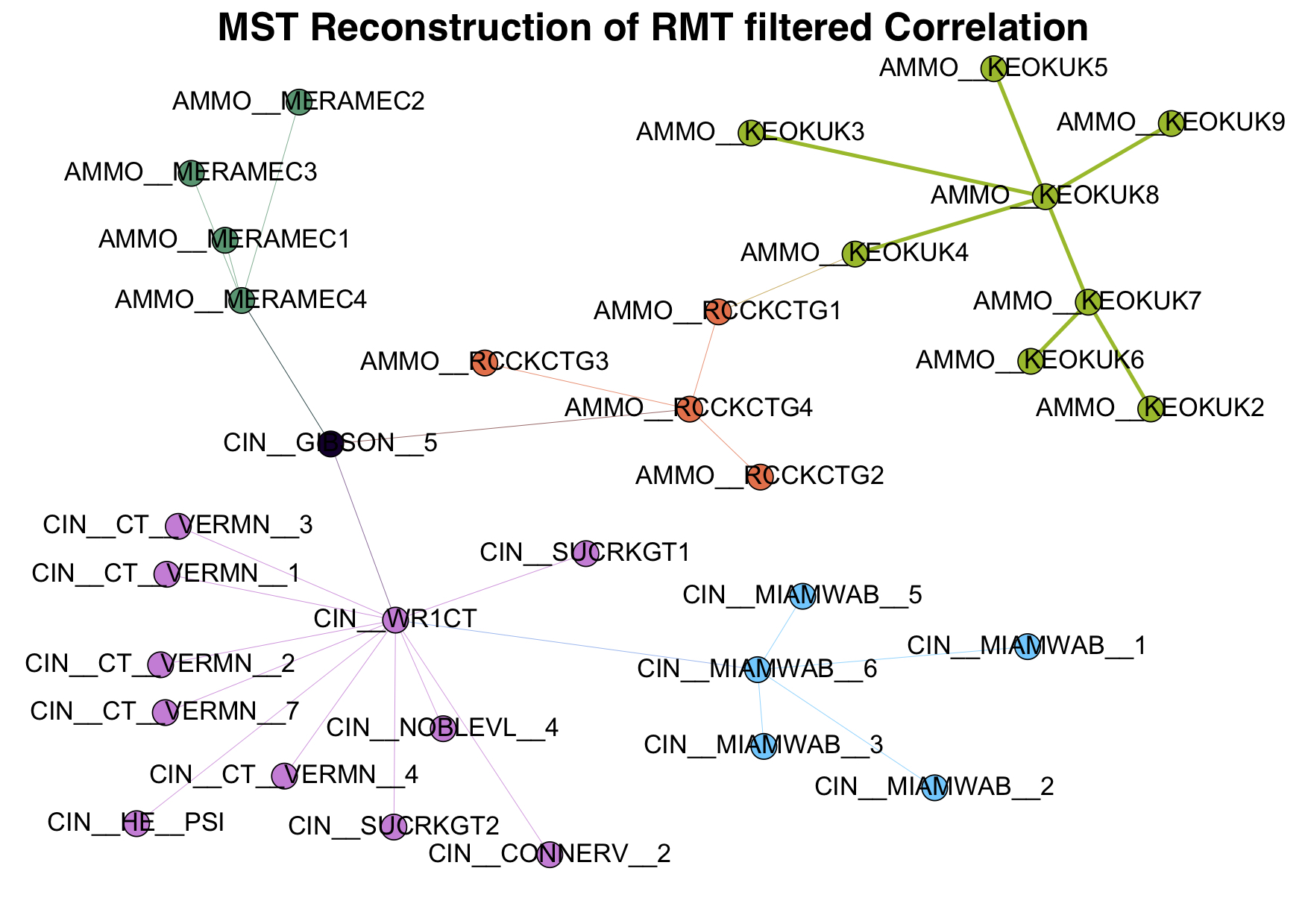}\\
	\includegraphics[scale=0.125]{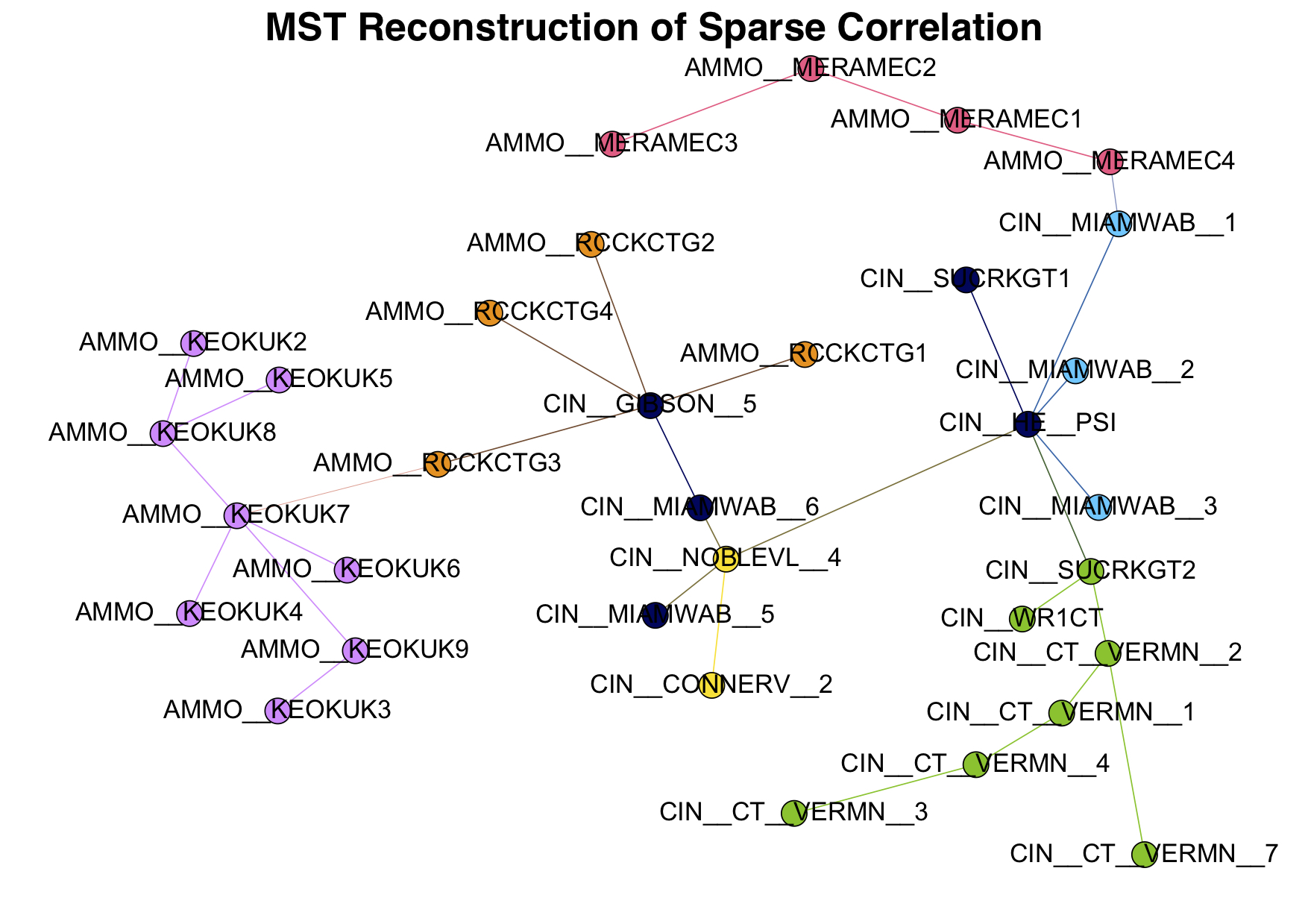}
	\includegraphics[scale=0.125]{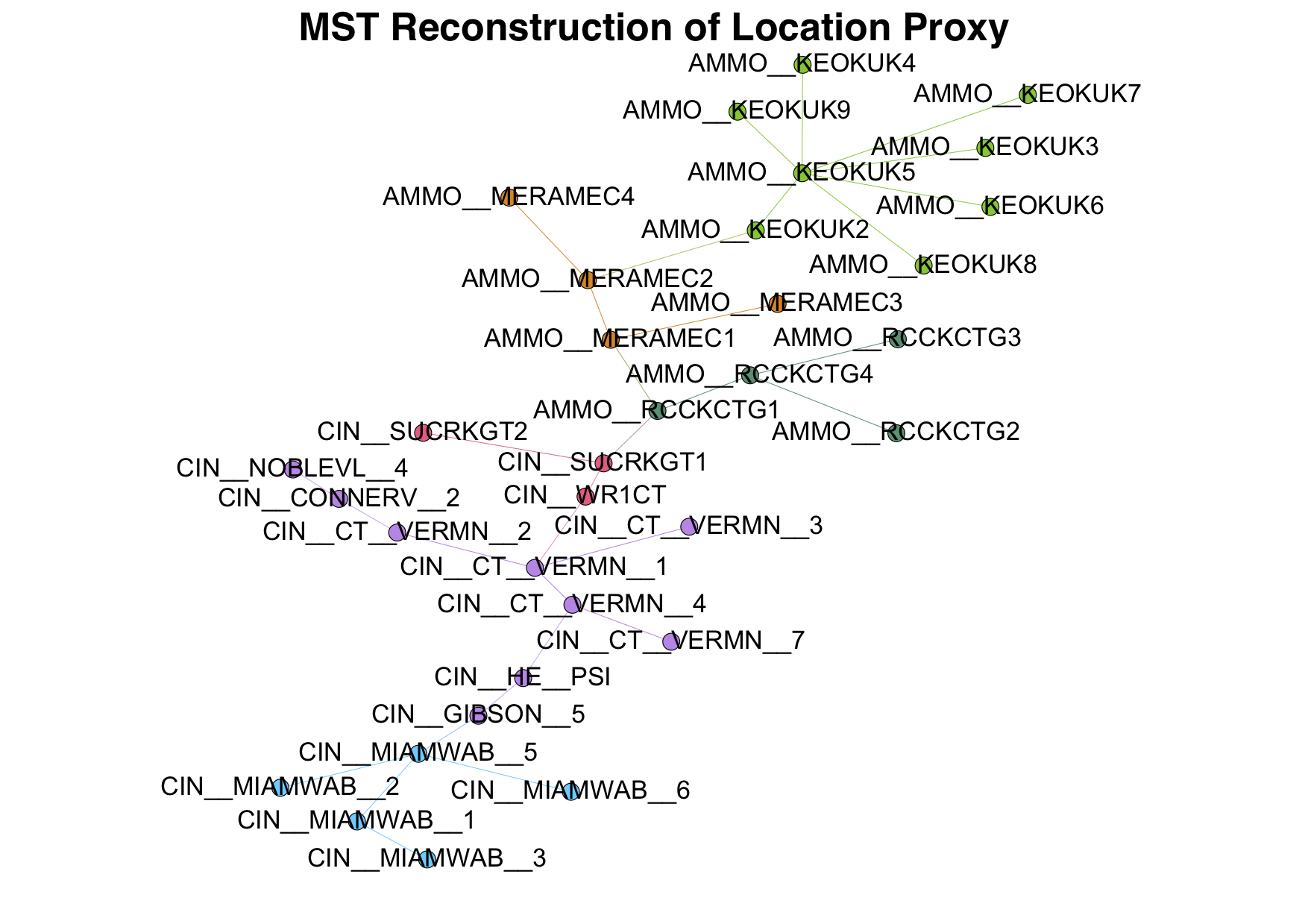}\\
	\caption{\textbf{Network reconstruction} for a small set of nodes of MISO using \textbf{MST filtering}.The location string of each node is shown in the figure. Nodes with same color belong to same cluster, which has been calculated by means \textbf{modularity maximization}. Black nodes are those misclassified with respect to location proxy partition. }
	\label{fig:MST result}
\end{figure*}

\begin{figure*}
	\includegraphics[scale=0.125]{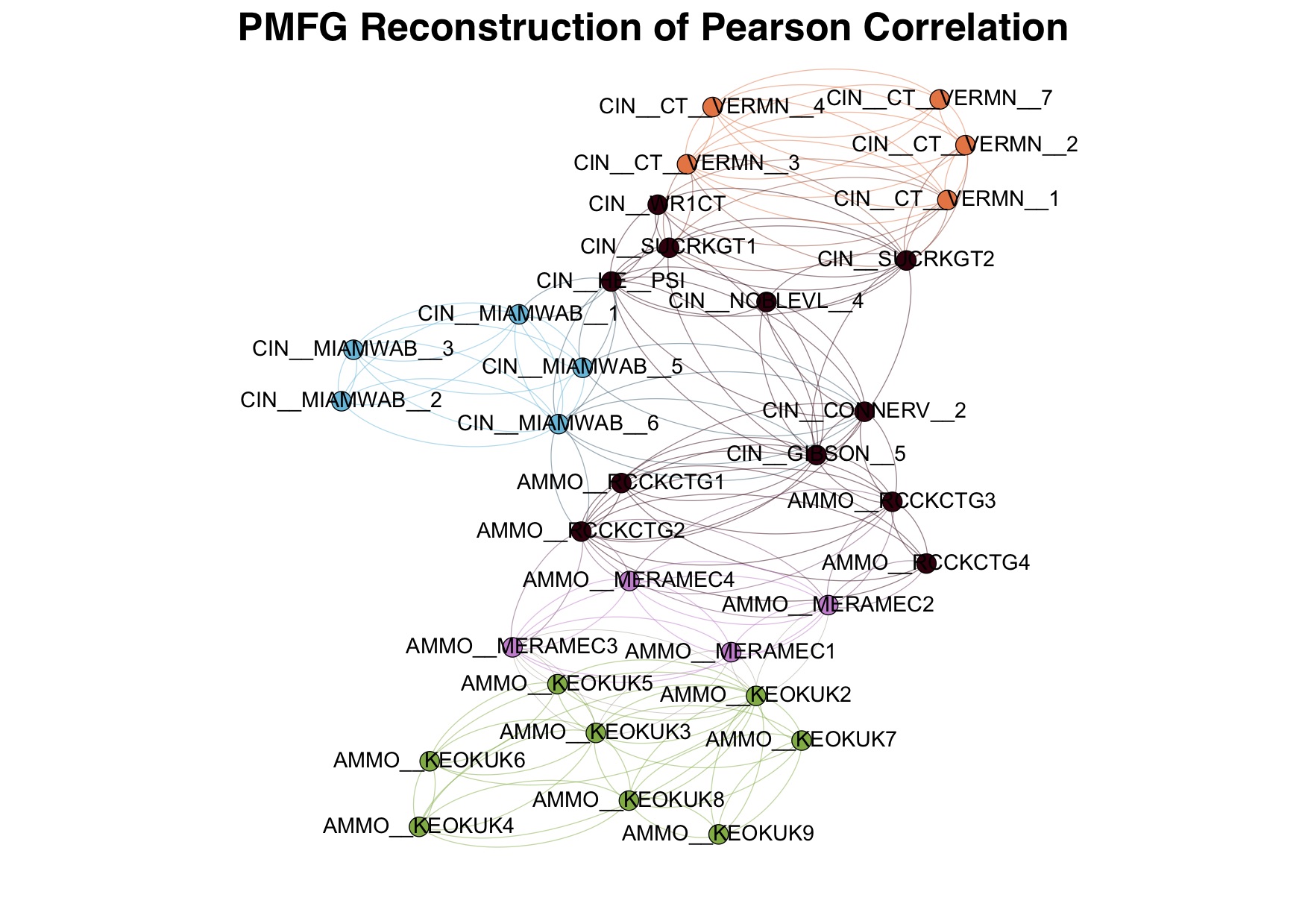}
	\includegraphics[scale=0.125]{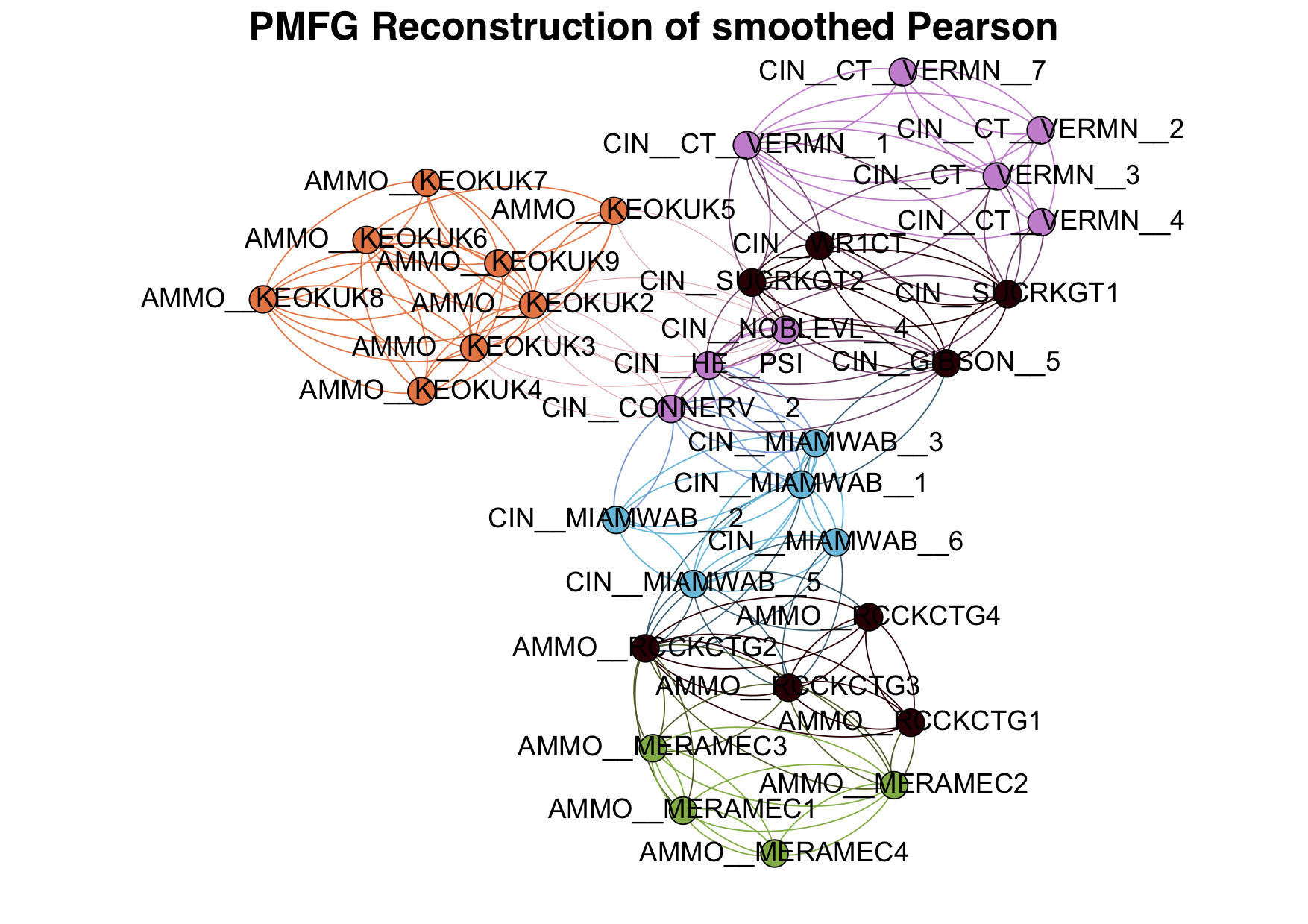}\\
	\includegraphics[scale=0.125]{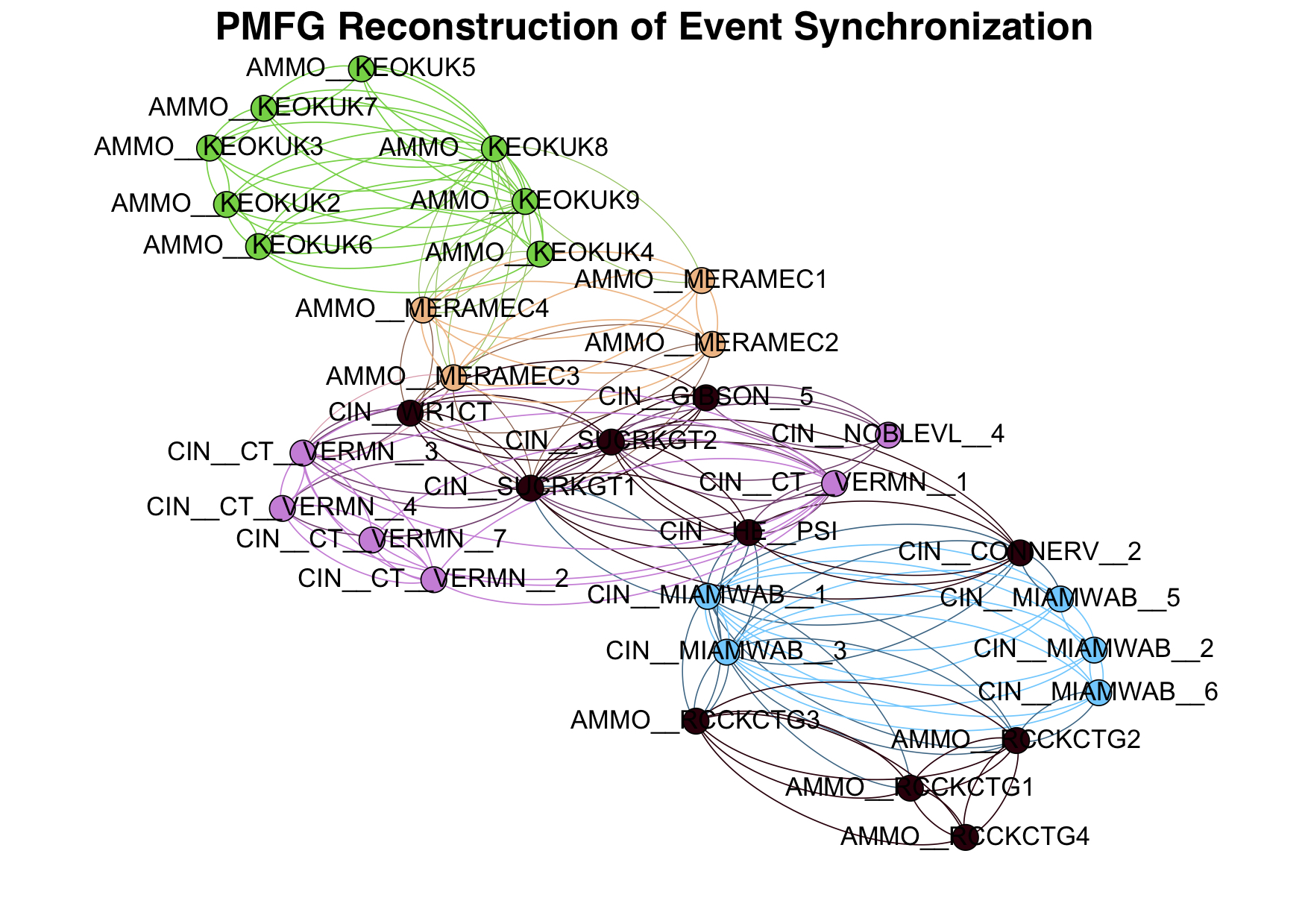}
	\includegraphics[scale=0.125]{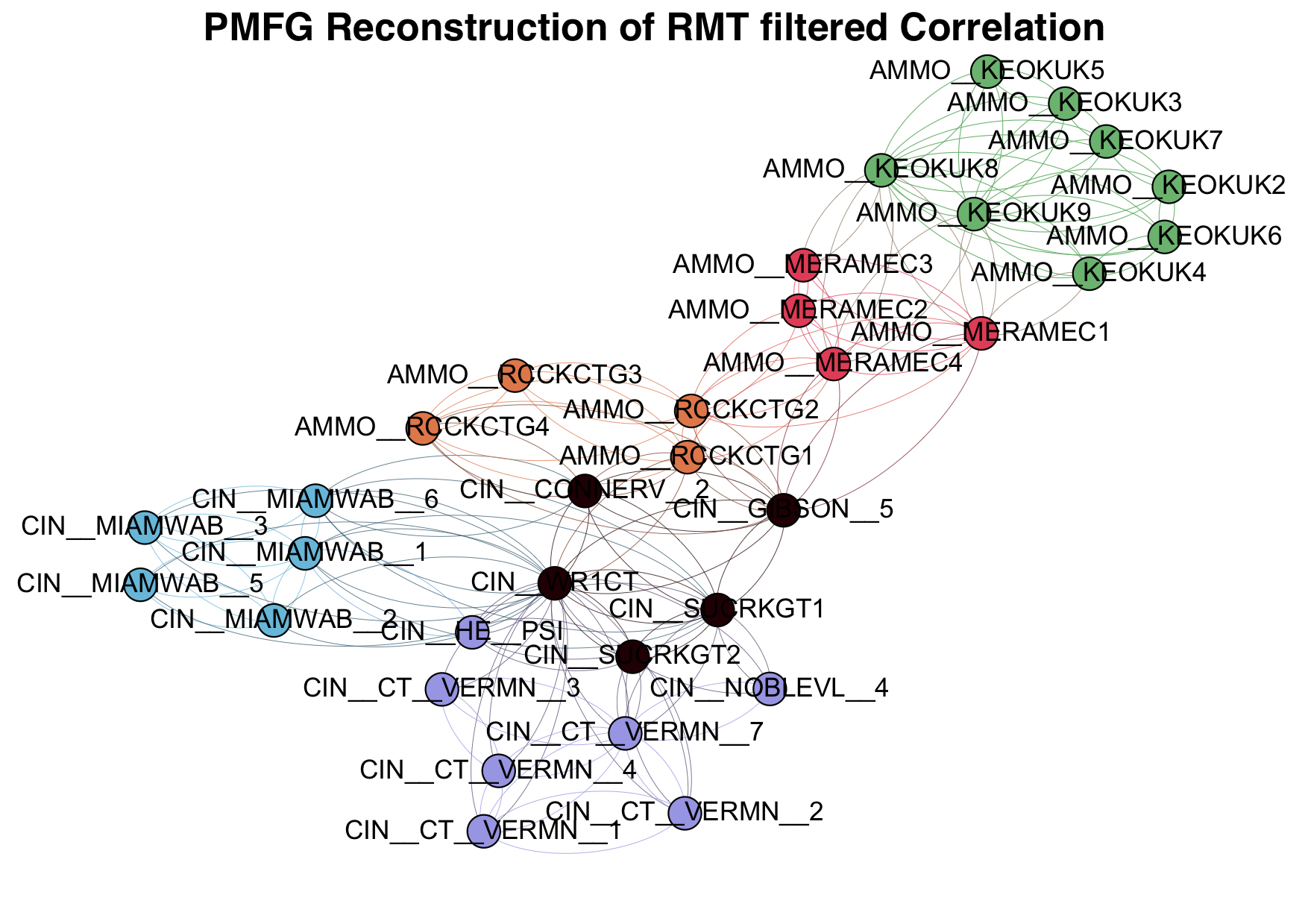}\\
	\includegraphics[scale=0.125]{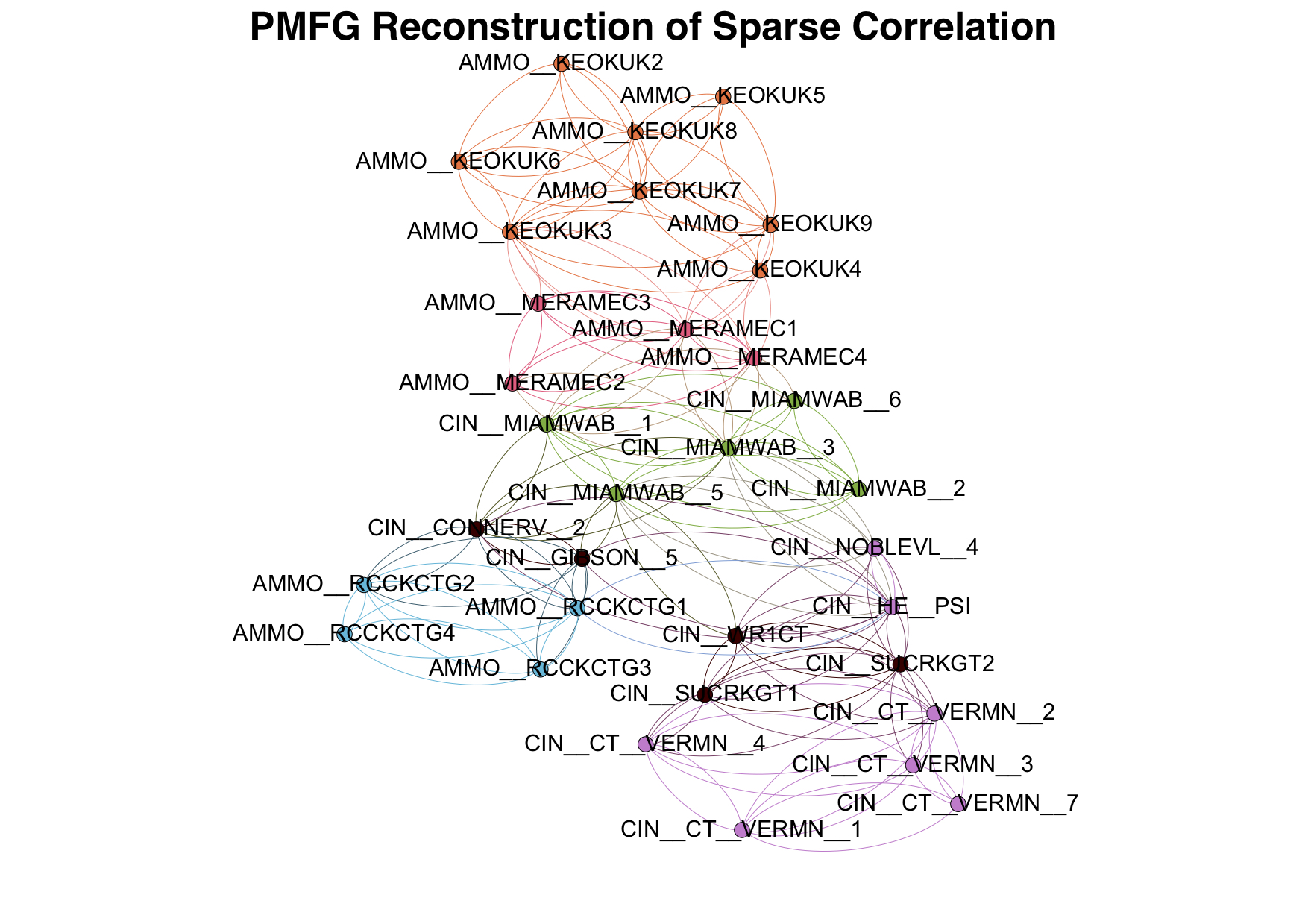}
	\includegraphics[scale=0.125]{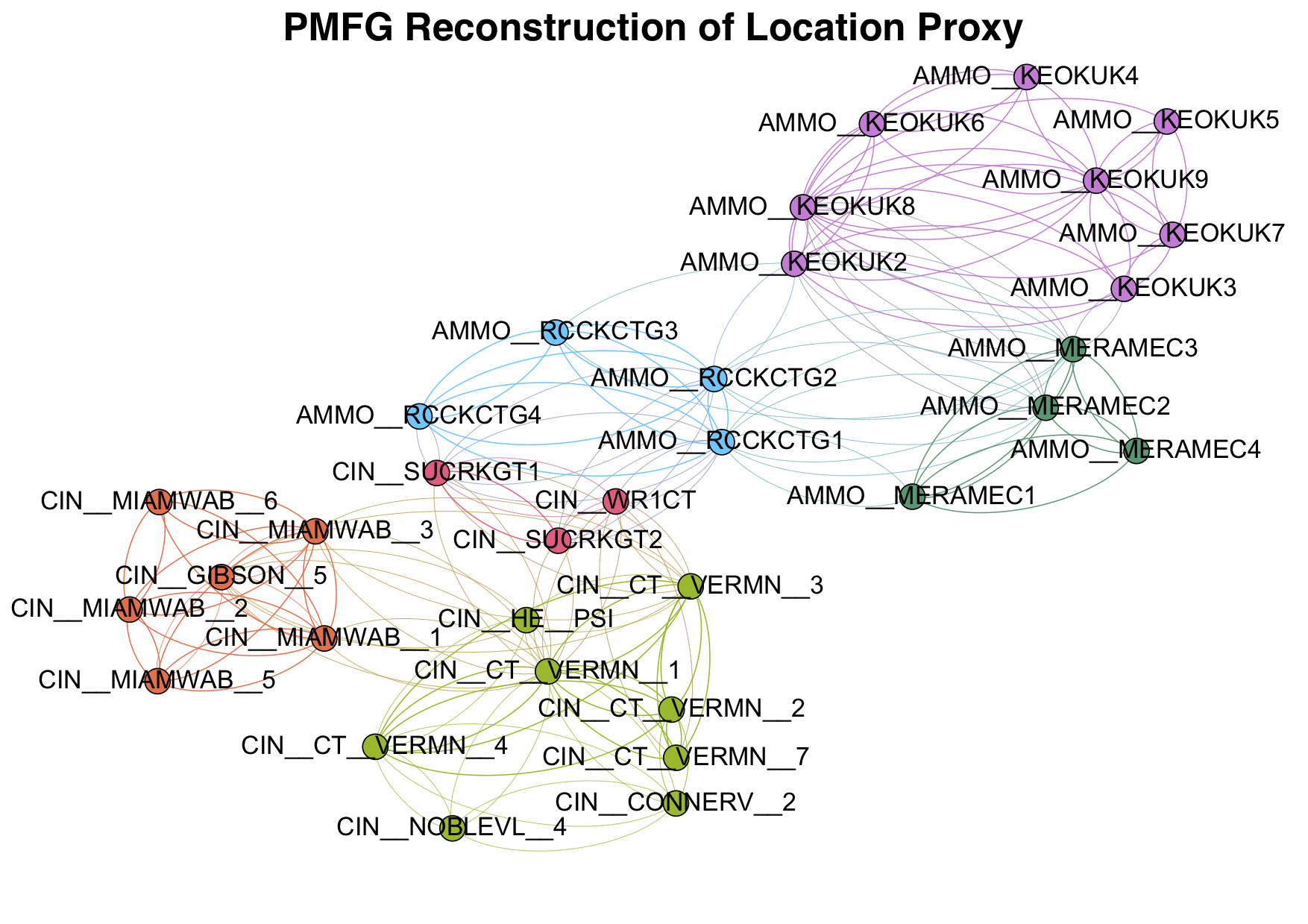}\\
	\caption{\textbf{Network reconstruction} for a small set of nodes of MISO with \textbf{PMFG filtering}. The location string of each node is shown in the figure. Nodes with same color belong to same cluster, which has been calculated by means of \textbf{modularity maximization}. Black nodes are those misclassified with respect to location proxy partition. }
	\label{fig:PMFG result}
\end{figure*}

As with the MST, the goal of this method to find the postulated approximating structure, rather than searching for groups of time series that are more correlated internally than with each other. Therefore we also need a graph community detection method, so we will use a modularity maximization algorithm to detect graph communities.
The modularity of a partition is a scalar value between $-1$ and $1$ that measures the density of links inside communities as compared to links between communities \cite{mm2}. In the case of weighted networks (weighted networks are networks that have weights on their links), it is defined as
\begin{equation}
Q=\frac{1}{m}\sum_{i,j}[A_{i,j}-\frac{k_{i}k_{j}}{2m}]\delta(c_{i},c_{j}),
\end{equation}
where $A_{i,j}$ represents the weight of the edge between $i$ and $j$, $k_{i} =  \sum_{j} A_{i,j}$ is the sum of
the weights of the edges attached to vertex $i$, and $c_{i}$ is the community to which vertex $i$ is assigned.
We use the algorithm proposed in \cite{mm}, which is fast and stable. See Fig. \ref{fig:MST result} for a small set of nodes of the clustering results based on the MST method, and Fig. \ref{fig:PMFG result} for PMFG.

In our toy example, we have two groups, CIN and AMMO, according to their locations. For AMMO, we have $3$ subgroups: MERAMEC, KEOKUK, and RCCKETG, which contains $4$, $8$, and $3$ elements respectively. For CIN, there are $3$ subgroups which contain more than $1$ elements, which are MIAMWEB ($5$ elements), VERMN ($5$ elements), and SUCRKGT ($2$ elements), and $5$ subgroups only contain $1$ element (we call them \textit{'single element'}): GIBSON, WR1CT, HE, CONNERV, and NOBLEVL. 

Ideally, a good partition should cluster elements from the same subgroup together, and if an element is single, it should be assigned to the similar subgroup. Thus, our location proxy is shown to provide a good benchmark (last figure in Fig.\ref{fig: Spectral clustering result}-\ref{fig:PMFG result}). By counting the black nodes, we find nearly all the methods can detect large subgroups (such as AMMO\_ KEOKUK and CIN\_ VERMN) correctly, and mistakes often occur when it comes to small subgroups. In our examples, the RMT filtered Correlation and Sparse Correlation perform better than others. We believe this to be due to the fact that these methods are more sensitive to smaller subgroups, where we find the other methods to provide less accurate results. As for clustering, we find that MST and PMFG are significantly better than spectral clustering, because they learn a sparse network structure which contains as much information as possible while filtering noise.

\section{Spectral clustering analysis}
In applying the spectral clustering method we must choose the number of clusters.  We will use the disparity (the heterogeneity measure) and Adjusted Rand Index (the similarity measure).  

\subsection{Clustering heterogeneity measure: disparity}
Disparity is used to measure the heterogeneity of cluster sizes. Some clustering methods might prefer many clusters with balanced size, while others might provide a subdivision into a few very large clusters and many small clusters \cite{musmeci2015relation}. In order to characterize the distribution of the cluster size with a single quantity, we can use the disparity measure \cite{musmeci2015relation}:
\begin{equation}
d = \frac{\sigma_{S}}{\bar{S}},
\label{disparity}
\end{equation}
where $\sigma_{S}$ is the standard deviation
\begin{equation}
\sigma_{S} = \sqrt{\frac{1}{N_{c}-1}\sum_{a}(S_{a}-\bar{S})^{2}}.
\label{cluster standard deviation}
\end{equation}
The normalization factor $\bar{S}$ is the average
\begin{equation}
\bar{S} = \frac{1}{N_{c}}\sum_{a}S_{a},
\label{average}
\end{equation}
where $S_{a}$ is the cardinality of cluster $a$ and $N_{c}$ is the number of clusters. The case where $d=0$ corresponds to all clusters being the same size, and large $d$ means that cluster sizes are more heterogeneous.

\subsection{Clustering similarity measure: Adjusted Rand Index}
The Adjusted Rand Index ($R_{ARI}$) is typically used to measure the similarity between the results of two clustering methods. In the case where we have the ground truth of the data partition, we can use $R_{ARI}$ as an accuracy measure for clustering algorithms.  If we assume that location provides a good partition for the LMP data, we can compare our clustering results by using $R_{ARI}$ with respect to the location proxy calculated from the string kernel correlation.
This method calculates the number of pairs of objects that are in the same cluster in both partitions, and then compares this with the one expected under the hypothesis of independent partitions \cite{musmeci2015relation}.
$R_{ARI}$ takes values in $[-1, 1]$, with $1$ corresponding to the case of identical clusters, and $0$ to two completely uncorrelated clusters \cite{musmeci2015relation}. 
Negative values show anti-correlation between two partitions, which means the number of overlapping pairs of these two partitions is less than the expected overlap between two random partitions \cite{musmeci2015relation}.

We calculate the ``contingency table'' $M$ between two partitions $Y$ and $Y'$ with coefficients \cite{musmeci2015relation}:
\begin{equation}
m_{ij} = |Y_{i}\cap Y'_{j}|.
\end{equation}
This is the number of objects in the intersection of clusters $Y_{i}$ and $Y'_{j}$. $M$ is a $k\times l$ matrix, where $k$ and $l$ are the number of clusters for $Y$ and $Y'$ respectively. Let us call $a$ the number of pairs of objects that are in the same cluster both in $Y$ and in $Y′$, and $b$ the number of pairs that are in two different clusters in both $Y$ and $Y′$ \cite{wagner2007comparing}. Then the Rand Index is defined as the sum of $a$ and $b$, normalized by the total number of pairs \cite{wagner2007comparing}:
\begin{equation}
R(Y, Y')=\frac{2(a+b)}{N(N-1)}=\sum_{i=1}^{k}\sum_{j=1}^{l}{m_{ij} \choose 2}.
\end{equation}
We then use as null hypothesis associated to two independent clusters the generalized hypergeometric distribution, and define the Adjusted Rand Index as the difference between the Rand Index and its mean value under the null hypothesis, normalized by the maximum that this difference can reach \cite{musmeci2015relation}:
\begin{equation}
R_{ARI} = \frac{R - t_{3}}{\frac{t_{1}+t_{2}}{2}-t_{3}},
\label{ARI}
\end{equation}
where $t_{1} = \sum_{i = 1}^{k}{|Y_{i}| \choose 2}$, $t_{2} = \sum_{j = 1}^{l}{|Y'_{j}| \choose 2}$, $t_{3} = \frac{2t_{1}t_{2}}{N(N - 1)}$.

We will use these two measures to tune the parameter in spectral clustering.

\subsection{Choice of the number of clusters}
In order to tune the number of clusters $n$ in spectral clustering we compare the clustering results with the location proxy partition for each type of correlation. The location proxy is the spectral clustering result based on the string correlation, which means we need to determine the number of clusters $n_{s}$ for the location proxy. This can be achieved by constructing a relation between $n$ and $n_{s}$:
\begin{eqnarray}
n_{s} =
\begin{cases}
144       & n \leq 144 \\
 n  & n > 144 
\end{cases},
\label{string information}
\end{eqnarray}
where $144$ is the number of different `name codes' in MISO (see section \ref{sec: string}). If $n_{s}=144$, we use exactly the same information as if we use the `code' partition. If $n_{s} > 144$, we use more information than the codes provide because some codes will be divided into several parts. 
If the clustering evaluation requires more information, it will grows as the number of clusters $n$ increases. We calculate $R_{ARI}$ with respect to the location proxy for each type of correlation measure with $n$ from $1$ to $300$. A good cluster number $n$ should make the clustering result stable, so we want to adjust the number of clusters until ARI is insensitive.
\begin{figure*}[h]
	\centering
	\includegraphics[scale=0.32]{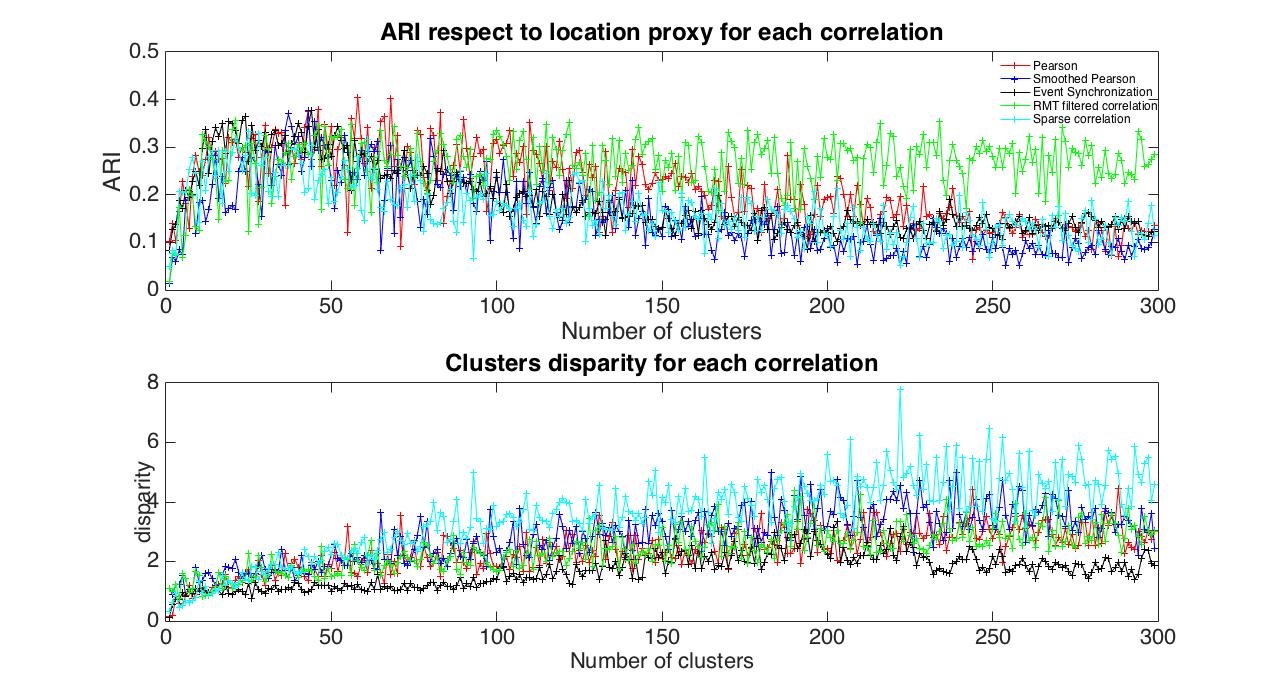}
	\caption{\textbf{Disparity and ARI as a function of the number of clusters based on five correlation measures.} Here we use the last $5000$ hours of LMP data for MISO. }
	\label{Tuning number of clusters}
\end{figure*}

Fig. \ref{Tuning number of clusters} shows how disparity and ARI change with respect to the number of clusters. As we can see, disparity and ARI converge when the number of clusters increasing. We can find, all of the ARIs increase at first, then decrease, and converge when the number of clusters are larger than (approximately) $200$. For RMT filtered correlations, the ARI converges to around $0.3$, while the others converge to $0.15$. As for disparity, these six correlations increase as the number of clusters grows, and converge to a value when the number of clusters reach about $200$.
Based on these result, we use $200$ as the number of clusters for spectral clustering.

\section{Dynamical analysis of the market structure}
In this section we analyze changes in the market structure as a function of time. We first analyze the behavior of the mean and the largest eigenvalues of each correlation matrix, then we analyze how the clustering results change through time.

The methodology we use has been introduced in \cite{musmeci2016multiplex} to estimate when structural changes (e.g. which span across the market) occur.

\subsection{Dynamical analysis of correlation matrices}
The mean of a correlation matrix represents the internal correlation strength of a set of time series, and the largest eigenvalue of an adjacency matrix provides information on the average degree of the graph\footnote{In fact, the largest eigenvalue $\lambda_{A}$ of the adjacency matrix $A$ of a general graph satisfies the inequality:
$\max(d_{avg},\sqrt{d_{max}})\leq \lambda_{A}\leq d_{max}$, where $d_{avg}$ is the average degree of nodes in the graph and $d_{max}$ is the largest degree.}. Also, it provides a sense of how each correlation is changing over time.
\begin{figure*}[h]
	\includegraphics[scale=0.32]{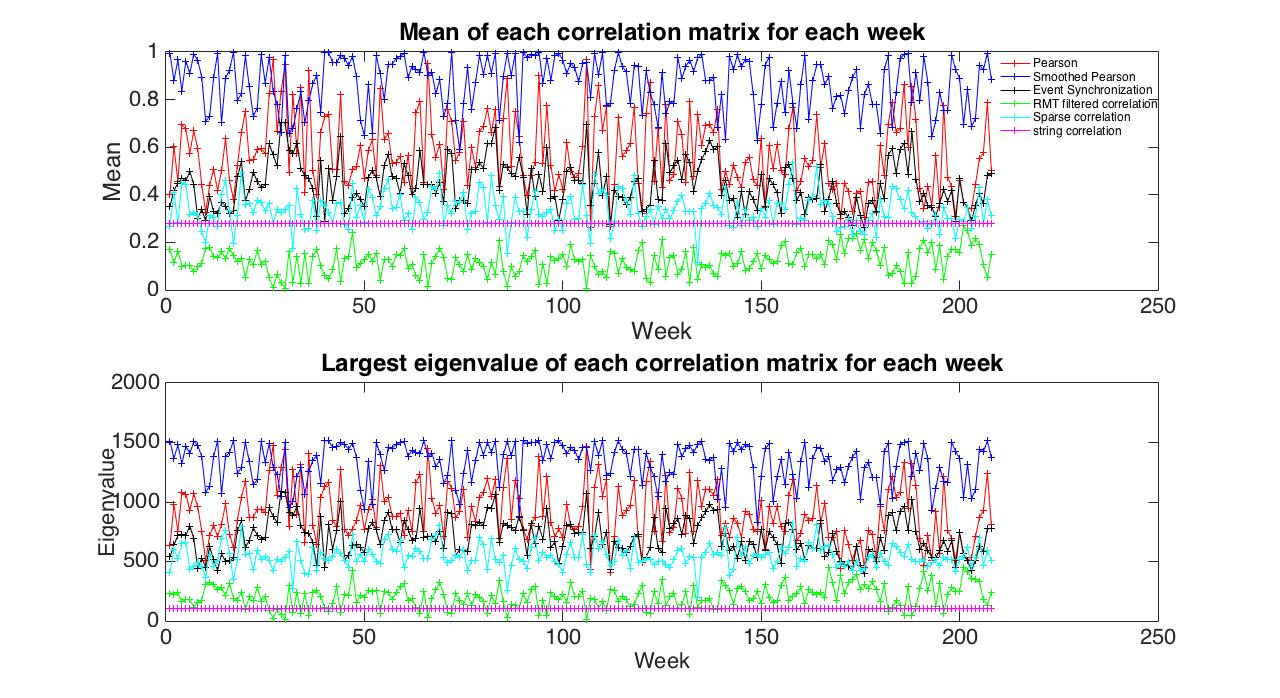}
	\caption{\textbf{Mean correlation and largest eigenvalue} of six different correlation measures for each week.}
	\label{mean and largest eigenvalue correlations}
\end{figure*} 
\begin{table}[]
	\footnotesize
	\centering

	\label{averaged mean and norm}
	\begin{tabular}{|c|c|c|}
		\hline
		& Mean   & Norm   \\ \hline
		Pearson 			             & 0.5703 & 905.46 \\ \hline
		smoothed Pearson & 0.8656 & 1329.5 \\ \hline
		Event synchronization            & 0.44   & 697.04 \\ \hline
		
		RMT filtered correlation         & 0.12   & 203.6  \\ \hline
		Sparse correlation               & 0.3323 & 539.84 \\ \hline
	\end{tabular}
		\caption{\textbf{Averaged mean and norm} for each correlation matrix}
\end{table}

Table \ref{averaged mean and norm} and Fig. \ref{mean and largest eigenvalue correlations} show the change of the mean and largest eigenvalue for different correlation matrices as a function of time. The RMT filtered correlation has the lowest correlation strength and highest sparsity, meanwhile the smoothed Pearson correlation has the highest strength and lowest sparsity.  We observe that Smoothed Pearson does change abruptly over time (both in mean and in the highest eigenvalue), but as we will see below, it is one of the two methods (together with Event Synchronization) which provides fairly stable ARI over time.



\subsection{Dynamical analysis of clustering results}
In this section we analyze the dynamical properties of the partitions using spectral clustering and Minimum Spanning Tree modularity. We consider $5$ different correlation measures\footnote{PMFG is not included because it is similar to MST but requires much more computational resources.}. 
We employed ARIs to measure the difference between two clustering results and disparity to measure the heterogeneity of cluster sizes. 
We considered two ARI measures: ``ARI location", which is based on the location proxy partition from the same clustering method, and ``ARI benchmark", which is based on the clustering from the corresponding correlation matrix of the last week (the $209$th week).

\subsubsection{Spectral clustering results analysis}
For each correlation method we calculate the correlation matrix of the LMPs for each week in the data, and use spectral clustering with a cluster number equal to $200$. 
\begin{table}[]
	\footnotesize
	\centering

	\label{clustering dynamic table}
	\begin{tabular}{|c|c|c|c|}
		\hline
		& Disparity & ARI\_benchmark & ARI\_location
		\\ \hline
		Pearson correlation              & 3.3031, 5.7603 (104)    & 0.1633, 0.3254 (97)         & 0.1303, 0.2256 (54)        \\ \hline
		Smoothed Pearson 			 & 3.8523, 9.426 (73)    & 0.0581, 0.1857 (76)         & 0.0595, 0.2255(108)         \\\hline
		Event synchronization            & 2.9588, 5.7923 (99)    & 0.1379, 0.3224 (3)         & 0.1203, 0.2355 (184)        \\ \hline

		RMT filtered correlation         & 8.6147, 12.2508 (118)    & 0.0716, 0.3633 (69)         & 0.0659, 0.386 (106)        \\ \hline
		Sparse correlation               & 6.866, 10.8894 (62)     & 0.1161, 0.5735 (69)         & 0.0411, 0.1709 (176)        \\ \hline
	\end{tabular}
		\caption{\textbf{Average and Maximum} disparity, ARI benchmark and ARI location of spectral clustering results for each correlation matrix. The number in parenthesis stands for the week when the maximum is reached.}
\end{table}
%
\begin{figure*}[h]
	\includegraphics[scale=0.32]{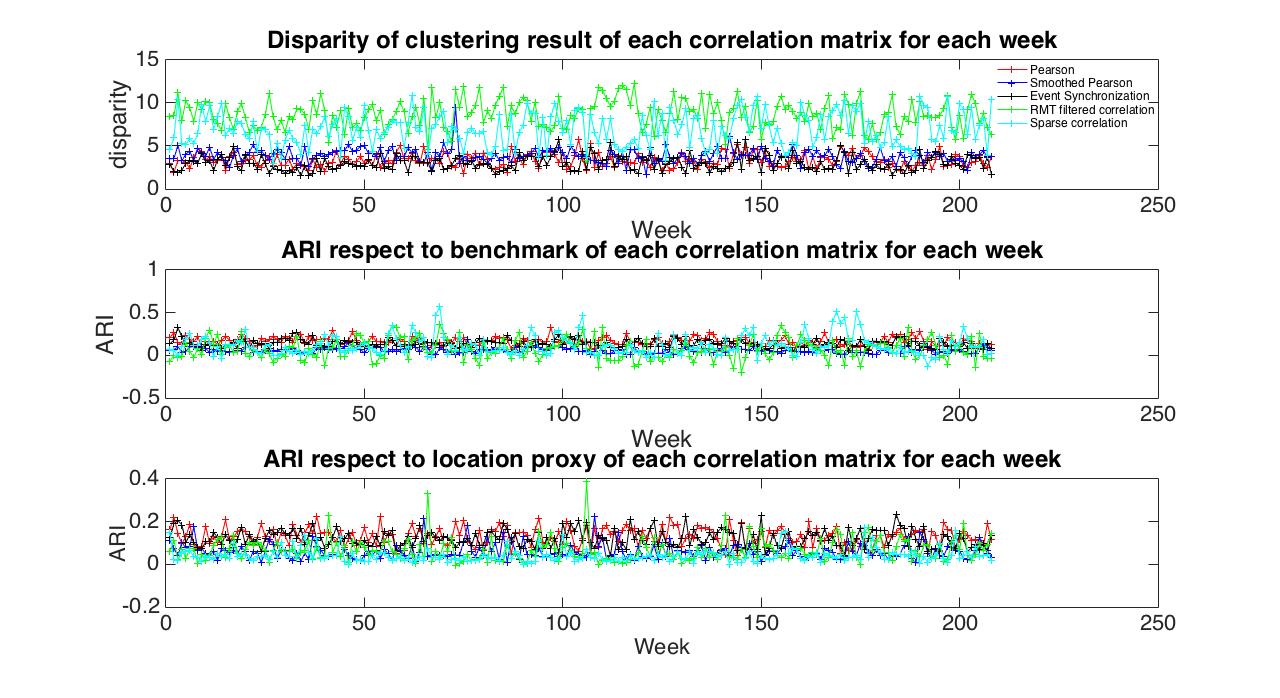}
	\caption{\textbf{The dynamics of disparities and AIRs of spectral clustering} results for $5$  correlation measures.}
	\label{spectral dynamics}
\end{figure*}

Figure \ref{spectral dynamics} shows the behavior in time of the spectral clustering results based on three different correlation measures. Table \ref{clustering dynamic table} shows the average and maximum\footnote{The number in brackets is the week that the measure reached its maximum.} of each spectral clustering.
We can see that disparity, in the case of RMT filtered correlations and the sparse correlation method are high with respect to the average (around $8.6$ and $6.8$ respectively). This means that the RMT filtered correlations and sparse correlation method prefer high heterogeneity of cluster sizes and less stable partitions in the spectral clustering setting.

The values attained by ARI with respect to the location proxy, smoothed correlation, RMT filtered correlation, and sparse correlation are respectively around $0.07$, while the latter are \textit{circa} $0.13$.  The reason is that these three methods remove a significant portion of the original correlation matrix, but the ARIs for both groups are quite small:  the difference is not significant in this case. The RMT filtered correlations have the highest maximum ARIs: this tells us it may have potential to out-perform other correlation methods. As for fluctuations, we see that all correlation measures have a similar behavior.  

%


\subsubsection{MST modularity maximization clustering analysis}
Similar with the analysis using spectral clustering, we construct a Minimum Spanning Tree for each correlation matrix, and find clusters based on the modularity maximization algorithms.
\begin{table}[]
	\footnotesize
	\centering

	\label{MST clustering dynamic table}
	\begin{tabular}{|c|c|c|c|}
		\hline
		& Disparity & ARI\_benchmark & ARI\_location
		\\ \hline
		Pearson correlation              & 0.3598, 0.4977 (144)    & 0.1114, 0.1941 (207)         & 0.0885, 0.1132 (24)        \\ \hline
		Smoothed correlation			 & 0.3758, 0.5221 (114)    & 0.0672, 0.1928 (208)         & 0.0697, 0.0952 (133)        \\ \hline
		Event synchronization            & 0.4236, 0.5736 (164)    & 0.1578, 0.3437 (208)         & 0.1413, 0.1847 (113)        \\ \hline

		RMT filtered correlation         & 1.0171, 2.5354 (56)     & 0.0772, 0.1223 (199)         & 0.2262, 0.3526 (117)        \\ \hline
		Sparse correlation               & 0.6527, 3.5332 (36)     & 0.1299, 0.3912 (207)         & 0.1576, 0.2988 (131)        \\ \hline
	\end{tabular}
\caption{\textbf{Average and Maximum} disparity, ARI benchmark and ARI location of \textbf{MST modularity maximization} results for each correlation matrix. The number in parenthesis stands for the week when the maximum is reached.}
\end{table}
%

\begin{figure*}[h]
	\includegraphics[scale=0.32]{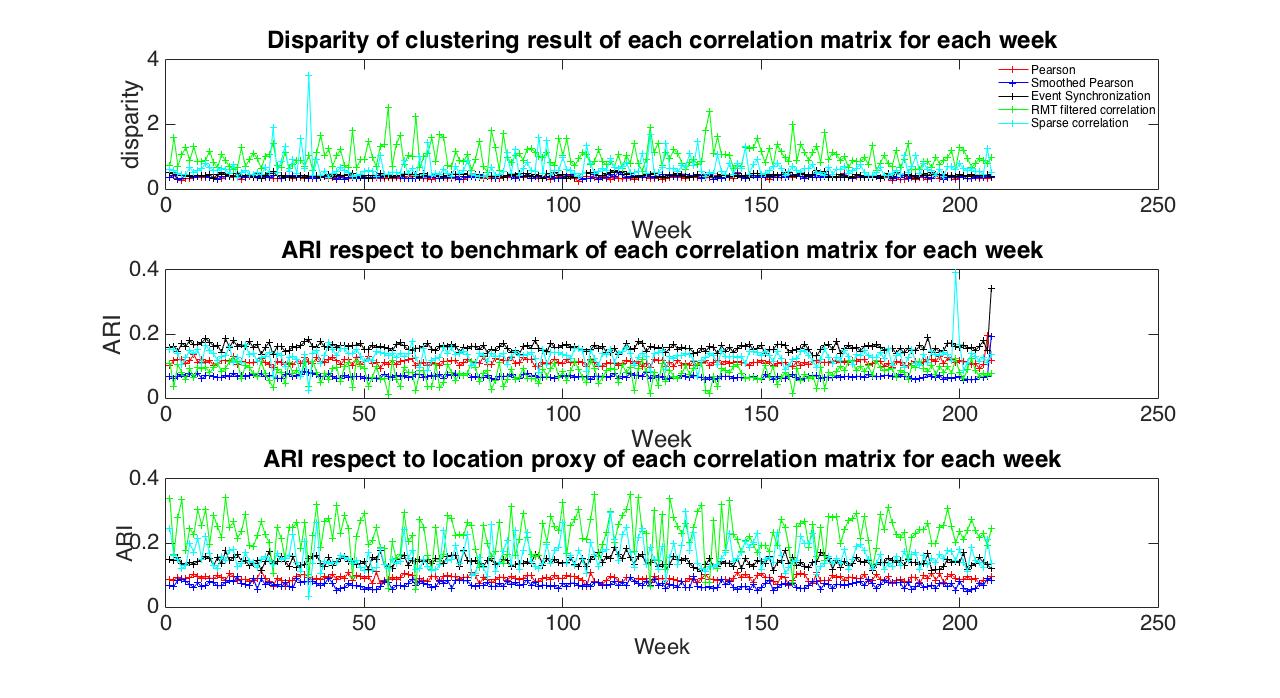}
	\caption{\textbf{The dynamics of disparities and AIRs of MST modularity maximization} for $5$ correlation measures. }
	\label{MST dynamics}
\end{figure*}

Figure \ref{MST dynamics} shows three measurements for the MST modularity maximization based on different correlation measures. Table \ref{MST clustering dynamic table} shows the average and maximum of each measurement.

The behaviour of dynamic disparity for MST clustering is similar to spectral clustering, while the RMT filtered correlations and sparse correlation measures have a higher average (around $1$ and $0.65$ respectively) and fluctuations of disparity, and the others have a similar average (around $0.38$) and fluctuations. 
For ARI with respect to the location proxy, event synchronization measure, RMT filtered correlations and sparse correlation measure are around $0.2$, while the others are about $0.07$. It can be concluded that using the MST modularity maximization method applied to these three correlations performance is significantly better than the Pearson and smoothed Pearson measures. 

From Figure \ref{spectral dynamics} and Figure \ref{MST dynamics} we further see that the disparity from MST is significantly lower than spectral clustering. The averaged ARI with respect to the location proxy is also higher than spectral clustering. Moreover, we find that the maximum ARI based on the benchmark is reached at around the $207$th week, which is very close to the benchmark ($209$th week), but the spectral clustering result in Table \ref{clustering dynamic table} does not show a similar behaviour. One possible explanation is that the spectral clustering method has more difficulty revealing patterns than MST modularity maximization in a large and dense network.

%
%
\begin{figure*}[h]
	\includegraphics[scale=0.33]{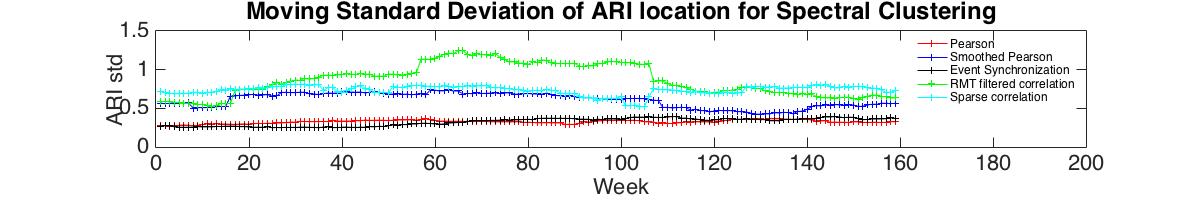}\\
	\includegraphics[scale=0.33]{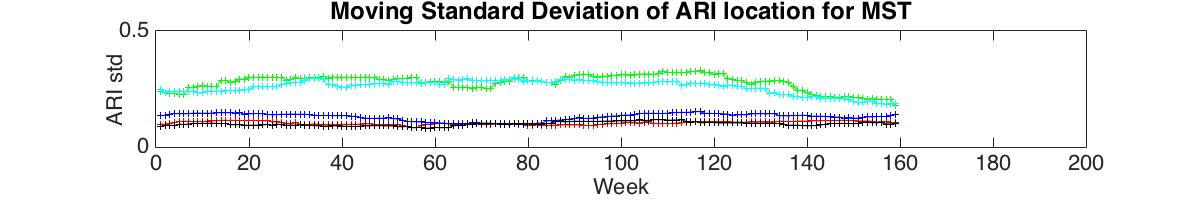}
	\caption{The moving standard deviation of ARI location for spectral clustering and MST modularity maximization with rolling window size equals to $50$. We find that MST provides more stable results than spectral clustering. RMT filtered correlation and sparse correlation are unstable in both situation.}
	\label{MSTD1}
\end{figure*}
From the moving standard deviation of ARI for spectral clustering and MST modularity maximization across time (Fig.\ref{MSTD1}), we observe that MST is significantly stable compared with spectral clustering. And RMT filtered correlation and Sparse Correlation are found to be unstable over time, while the other correlation methods can be considered stable.
This suggests that the electricity market structure revealed by RMT and MST modularity maximization changes frequently in time and is thus seems not reliable for these markets. Of course, here a note of warning should be made: this depends on our proxy measure, which is the string correlation method. In general, we believe this to be due to the properties of the market itself, and the strong non-stationarity, on which these methods strongly rely upon. However, methods based on Pearson and Event Synchronization, in particular MST, were found to be rather stable. It should be said, however, that even in the case of financial market time series, previous findings \cite{Dong2005} show that there are only a few clusters in stock markets which are stable over time (using RMT filtered correlation and modularity maximization). This suggests that electricity markets are far more unstable than other financial markets. One possible explanation of the non-stationarity is that electricity markets are based on a strongly non-local optimization problem which spreads disturbances and creates correlations across the entire footprint of the grid. Nonetheless, the method based on Event Synchronization (which focuses only on rare and strong events) and (Smoothed and non) Pearson are less sensitive to fluctuations.

\section{Conclusions}
In this paper we discussed various correlation, filtering, and clustering methods with the purpose of analyzing the underlying structure of wholesale electricity market data. Specifically, we have analyzed the nodal price difference between day-ahead and real-time, which are relevant for understanding the inefficiencies in the planning process that outputs the day-ahead prices.
We also introduced an alternative measure of correlation based on a modification of Event Synchronization, which focuses on more rare and strong events (spikes) in time series. We have argued that most of the correlations observed in our dataset are due to random fluctuations and that these should be filtered, or a nonlinear measure of correlation should be used instead.
In addition to this, we explored various methods for inferring the correlation structure for MISO, identifying various clusters. These methods are robust as they use only the strongest correlation links.

Finally and most importantly, we provided a framework to choose and filter correlation matrices of electricity price data by using the string kernel correlation proxy, which we believe can be used to test the validity of our methodology. 

We also extended our results to the time domain and analyzed how the market structure changes from week to week. This analysis provides a quantitative method to infer non-stationarity as well as to test the clustering methods. We concluded that the MST modularity maximization method is more suitable than spectral clustering.
Furthermore, even after filtering multiple types of noise, the clustering results for RMT filtered correlation matrices were found to be, in some cases, unstable over time (with respect to the proxy), although it has achieved the best performance. Of all the methods, we have found Pearson based and the Event Synchronization methods to be the most stable overtime. Overall, we have found Event Synchronization in conjunction with MST to be the most appropriate clustering method for its stability and remarkable performance.
An interesting avenue for future work is to try to reconstruct the network using methods of graph inference based on entropy maximization \cite{tiziano}.\ \\\ \\


\begin{centering}
\textbf{Acknowledgements}\\
\end{centering}
FC and TC would like to thank Invenia Labs for previous funding and support respectively. We would like to thank L. Roald at LANL for comments on an earlier version of this paper. FC acknowledges the support of NNSA for the U.S. DoE at LANL under Contract No. DE-AC52-06NA25396. TC acknoledges the support of National scientific and Technological Innovation Zone (No. 17-H863-01-ZT-005-005-01) and State's Key Project of Research and Development Plan (No. 2016QY03D0505). \ \\\ \\

\begin{centering}
\textbf{Bibliography}\ \\\ \\
\end{centering}

\end{document}